\documentstyle[12pt]{article}

 \catcode`\@=11
\def\Bbb{\ifmmode\let\next\Bbb@\else
 \def\next{\errmessage{Use \string\Bbb\space only in math mode}}\fi\next}
\def\Bbb@#1{{\Bbb@@{#1}}}
\def\Bbb@@#1{\fam\msbfam#1}

\if@twoside
\else
\fi
\ifcase \@ptsize
\or % mods for 11 pt
\or % mods for 12 pt
\fi

\def\@citex[#1]#2{%
\if@filesw \immediate \write \@auxout {\string \citation {#2}}\fi
\@tempcntb\m@ne \let\@h@ld\relax \def\@citea{}%
\@cite{%
  \@for \@citeb:=#2\do {%
    \@ifundefined {b@\@citeb}%
      {\@h@ld\@citea\@tempcntb\m@ne{\bf ?}%
      \@warning {Citation `\@citeb ' on page \thepage \space undefined}}%
      {\@tempcnta\@tempcntb \advance\@tempcnta\@ne%
      \@tempcntb\number\csname b@\@citeb \endcsname \relax%
      \ifnum\@tempcnta=\@tempcntb %   Number follows previous--hold on to it
        \ifx\@h@ld\relax%
          \edef \@h@ld{\@citea\csname b@\@citeb\endcsname}%
        \else%
          \edef\@h@ld{\ifmmode{-}\else--\fi\csname b@\@citeb\endcsname}%
        \fi%
      \else%   %  non-successor--dump what's held and do this one
        \@h@ld\@citea\csname b@\@citeb \endcsname%
        \let\@h@ld\relax%
      \fi}%
    \def\@citea{,\penalty\@highpenalty\,}%
  }\@h@ld
}{#1}}
\@addtoreset{equation}{section}
\def\section{\@startsection {section}{1}{\z@}{-3.5ex plus -1ex minus
 -.2ex}{2.3ex plus .2ex}{\large\bf\centering}}
\def\subsection{\@startsection{subsection}{2}{\z@}{-3.25ex plus -1ex minus
 -.2ex}{1.5ex plus .2ex}{\sc}}
\gdef\@publabel{\hfil}
\gdef\@pubdate{\null}
\gdef\@pubnumber{\null}
\gdef\@author{\null}
\gdef\@title{\null}
\gdef\@abstract{\null}
\long\def\pubdate#1{\gdef\@pubdate{#1}}
\long\def\pubnumber#1{\gdef\@pubnumber{#1}}
\long\def\publabel#1{\gdef\@publabel{#1}}
\long\def\author#1{\gdef\@author{#1}}
\long\def\title#1{\gdef\@title{#1}}
\long\def\abstract#1{\gdef\@abstract{#1}}
\def\titlerelax{
\let\maketitle\relax
\let\settitleparameters\relax
\let\consolidatetitle\relax
\let\inittitlepage\relax
\let\finishtitlepage\relax
\let\titlepagecontents\relax
\let\multithanks\relax
\let\titlebaselines\relax
\let\@makepub\relax
\let\@maketitle\relax
\let\@makeauthor\relax
\let\@makeabstract\relax
\let\@maketitlenote\relax
\let\thanks\relax
\let\titlerelax\relax}
\def\titleclean
{\gdef\@titlenote{}
\gdef\@abstract{}
\gdef\@author{}
\gdef\@title{}
\gdef\@pubdate{}\gdef\@pubnumber{}\gdef\@publabel{}
\gdef\@dpublabel{}
}
\def\@makepub{\vbox to \z@{\hbox to \textwidth{\hfill
\@publabel \hfill
\llap{\parbox[t]{0.25\textwidth}{\raggedleft\@pubnumber}}}%
\vss}}
\def\@maketitle{\vskip 60pt \begin{center}
 {\LARGE \@title \par}
 \end{center}}
\def\@makeauthor{{%
\def\and{\smallskip {\normalsize \rm and\smallskip }}
\def\And{\medskip {\normalsize \rm and\\}\medskip}
\long\def\address##1{{\def\and{\\and\\}\medskip
				{\small \it \\##1\\}
}}
{\centering
 \vskip 3em
 \large \lineskip .75em
 \@author}
 \par}}
\def\@makedate{\vskip 1.5em
 {\raggedright \small \noindent\@pubdate \par}}
\def\@makeabstract{\vskip 1.5em
{\small
\begin{center}
{\bf ABSTRACT\vspace{-.5em}\vspace{0pt}}
\end{center}
\quotation \@abstract \endquotation}}
\def\maketitle{\titlepage
\let\footnotesize\small \setcounter{page}{0}
\@makepub
\vfil
\@maketitle
\@makeauthor
\vfil
\@makeabstract
\@thanks
\vfil
\@makedate
\if@restonecol\twocolumn \else \eject \fi
\titlerelax \titleclean
\setcounter{footnote}{0}
}

\catcode`\@=12

\begin{document}
\bibliographystyle{npb}

%b
\let\b=\beta
\def\blank#1{}

%c
\def\cdd{{\cdot}}
\def\cev#1{\langle #1 \vert}
\def\cH{{\cal H}}
\def\comm#1#2{\bigl [ #1 , #2 \bigl ] }
\def\compact{ reductive}
\def\cont{\nonumber\\*&&\mbox{}}
\def\cO{{\cal O}}
\def\cul #1,#2,#3,#4,#5,#6.{\left\{ \matrix{#1&#2&#3\cr #4&#5&#6} \right\}}

%d
\def\dz{Dz}
\def\dz{\hbox{$d\kern-1.1ex{\raise 3.5pt\hbox{$-$}}\!\!z$}}
\def\dz{ \frac{d\!z}{2\pi i}}
%e
\def\en{\end{equation}}
\def\enn{\end{eqnarray}}
\def\eq{\begin{equation}}
\def\eqq{\begin{eqnarray}}

%f

%g

%h
\def\half#1{\frac {#1}{2}}

%i
\def\ip#1#2{\langle #1,#2\rangle}

%j

%k
\def\k{k}

%l

%m
\def\Mf#1{{M{}^{{}_{#1}}}}
\def\mno{{\textstyle {\circ\atop\circ}}}
\def\mod#1{\vert #1 \vert}

%n
\def\Nf#1{{N{}^{{}_{#1}}}}
\def\ni{\noindent}
\def\no{{\textstyle {\times\atop\times}}}
\def\no:#1:{\mno#1\mno}
\def\nox{{\scriptstyle{\times \atop \times}}}

%o

%p
\let\p=\phi
\def\posdef{ positive-definite}
\def\posdefness{ positive-definiteness}
%q
\def\Qf#1{{Q{}^{{}_{#1}}}}
\def\Qstar{\mathop{\no:QQ:}\nolimits}

%r
\def\reductive#1#2{#1}

%s

%t
\def\tr{\mathop{\rm tr}\nolimits}
\def\Tr{\mathop{\rm Tr}\nolimits}

%u

%v

%r

%s

%t

%u

%v
\def\vec#1{\vert #1 \rangle}
\def\vac{\vec 0}

%w
\def\wan{$\WA_n$ }
\def\Wb{\bar W}
\def\Wf#1{{W{}^{{}_{#1}}}}
\def\wbn{$\WB_n$ }
\def\WA{\mathop{\it WA}\nolimits}
\def\WB{\mathop{\it WB}\nolimits}
\def\WBC{\mathop{\it WBC}\nolimits}
\def\WD{\mathop{\it WD}\nolimits}
\def\WG{\mathop{\it WG}\nolimits}

%x

%y

%z
\def\zz#1{(z-z')^{#1}}

\openup 1\jot

\pubnumber{ DTP 94-5}
\pubdate{Mar. 1994}

\title{ Representation Theory of a W-algebra from generalised DS reduction}

\author{
P. BOWCOCK\thanks{ Email \tt
Peter.Bowcock@durham.ac.uk}
\address{Dept. of Mathematical Sciences,
University of Durham,
Durham, DH1 3LE, U.K.}
}

\abstract{We investigate the W-algebra resulting from Drinfel'd-
Sokolov reduction of a $B_2$ WZW model with respect to the grading induced by a
short
root. The quantum algebra, which is generated by three fields of spin-2
and a field of spin-1, is explicitly constructed. A `free field'
realisation of the algebra in terms of the zero-grade currents is
given, and it is shown that these commute with a screening charge. We
investigate the representation theory of the algebra using a
combination of the explicit fusion method
of Bauer et al. and free field methods. We discuss the fusion rules of
degenerate primary fields, and give various character formulae and a
Kac determinant formula for the algebra
}

\maketitle

\section{Introduction}

Perhaps the outstanding goal in conformal field theory \cite{BPZ} is a complete
classification. At present such a classification seems to remain beyond
our scope. We may hope that if we study enough examples of conformal field
theory, the patterns that we discover may provide us with clues as to
how we should go about a classification.

Until recently, the most general construction of rational conformal field
theories available was the coset construction of Goddard, Kent and Olive
\cite {GKOl}.
This construction has many appealing features; in particular,
one can write down the partition function and fusion rules of the theory
straightforwardly. On the other hand, the symmetry, or chiral algebra,
of the theory seems somewhat obscured in this formulation \cite{BGod,Doug}.
For instance,
it is not a straightforward matter to see that the coset models
${\widehat su}(2)_2\times {\widehat su}(2)_k/{\widehat su}_{k+2}$
are in fact superconformal minimal models \cite {GKOl}, or that the
models $\hat g_1\times \hat g_k/\hat g_{k+1}$ are minimal models for the
algebras $W_g$ \cite{BBSS,Watt3}.

In the last few years another construction of conformal field theory has
arisen.
This is based on
a generalised version of the Drinfel'd-Sokolov reduction
\cite{DSok,BOog,ASha,Bela,GERV,BFFOW1}.
Many of the features of the
resulting W-algebras can be easily derived from simple finite Lie-algebraic
concepts \cite{Int,BTvD}, and it has been argued using cohomological
techniques that the
algebras can always be quantised \cite{FFre3,BTji}.
However, as yet surprisingly little is known about the representation
theory of this new class of algebras, except for the
algebras $W_g$ which correspond to the standard reduction of a WZW model
based on a simply-laced Lie algebra $g$ \cite{FLuk,FLuk2,FKWa}.
As a result, the primary field content,
fusion rules and partition function of the conformal field theories are poorly
understood.

In this paper, we shall try to further our understanding of
the representation theory of W-algebras
constructed from generalised Drinfel'd-Sokolov reductions by examining one
particular example of a such a W-algebra in rather thorough detail.
Our example, which we shall hereafter refer to as $\Wb$, is based on the
reduction of $B_2$ by the nilpotent algebra associated
with the grading induced by the short root of $B_2$. This algebra is a
relatively simple object generated by three spin-2 fields and a single spin-1
field, yet it contains all of the important features that we expect the general
case to have.
The main tool that we shall use to analyze the representation theory
of this algebra is the explicit fusion technique pioneered in
\cite{BDIZ1,BPTa1}.

The structure of the paper is as follows.
In section 2 we give a quick review of generalised Drinfel'd-Sokolov reduction.
In the third section we construct the quantum version of our example.
In addition we quantise the Miura transformation and give
a `free field' construction of the algebra in terms of the {\it non-abelian}
zero grade currents of $B_2$. These expressions are shown to commute with the
appropriate screening charge.

In section 4 we begin our investigation of the representation theory by
examining the zero-grade algebra, which acting on highest-weight states
is a finite W-algebra in the terminology of \cite{Tjin1}. Coincidentally this
algebra has arisen before in the literature. We then use a number of
low-lying null vectors in section 5 to
explicitly derive the fusion of the basic primary fields of the theory
with other fields using the methods of Bauer et al. and Bajnok et al..
This also gives us a recurrence relation which can be used to derive
explicit formulae for a subset of all the possible null vectors.

Combining this information with what we might expect from the free
field form of the algebra and our experience with other W-algebras,
in section 6 we conjecture the general
form of degenerate representations, the fusions of the corresponding
fields, formulae for the characters of the representations and the
Kac determinant formula for this algebra. We conclude with some
comments on what we have found and directions for future research.

\section{Review of the generalised DS reduction}

\label{sec.hs}
\label{sec.u10}
\def\rhov{{\rho^\vee}}
\def\lambdav{{\lambda^\vee}}

Both the coset construction and the method of Hamiltonian reduction can be
thought of as a gauging of the WZW model \cite{Witt1} whose action is given by
\eq
S(u)={\hbar k\over 4}\left [{1\over 2}\int d^2x
{\rm Tr} (\partial^\mu u \partial_\mu u^{-1})
+{1\over 3}\int d^3x {\rm Tr} (\epsilon^{ijk}u^{-1}\partial_i u u^{-1}
\partial_j u u^{-1}\partial_k u)\right ]
\label{eq.wzw}
\en
where the field $u$ takes its value in the group $G$ with Lie algebra $g$.
When quantised, the modes of the left and right currents
\eq
J_L(x_+) =  {\hbar k\over 4}u^{-1}\partial_+ u\,,\,
J_R(x_-) = {\hbar k\over 4} \partial_- uu^{-1}
\label{eq.currents}
\en
of the WZW model form two
commuting copies of a Kac-Moody algebra ${\hat g}$
\eq
{}~[T^a_m,T^b_n]=f^{ab}_c T^c_{m+n}+ km\delta_{m+n,0}.
\label{eq.kma}
\en
where $f^{ab}_c$ are the structure constants of the algebra $g$.
By Noether's theorem, $J_L$, $J_R$ generate the transformations
$u\to uv(x_+)$, $u\to w(x_-)u$ respectively where $v,w$ are elements of $G$.
In the coset construction one gauges a vector subgroup of the symmetry of the
WZW model by adding a number of terms to the action \cite{FSon}
\eqq
S(u,A_R,A_L) &=& S(u) +\int d^2x {\rm Tr} (A_LJ_R-A_RJ_L)\cont
+{\hbar k\over 4}\int d^2x {\rm Tr} (A_LuA_Ru^{-1}-A_LA_R)
\label{eq.gwzw}
\enn
where $A_L,A_R\in {\hat h} \subset {\hat g}$.

Now let us turn our attention to DS reductions. These also are a gauging of
the WZW model, but in this case one gauges the currents associated with
nilpotent directions of the algebra. This can be achieved by
altering the action ({\ref{eq.gwzw}) to
\eq
S(u,A_R,A_L) = S(u) +\int d^2x {\rm Tr}
[A_L(J_R-I_-)-A_R(J_L-I_+)+ {{\hbar k}\over 4}A_LuA_Ru^{-1}],
\label{eq.gwzwds}
\en
restricting $A_\pm$ to
nilpotent directions of the algebra \cite{BFFOW1}
and taking $I_\pm$  to be certain constant elements of $g$ which we will
discuss
below. Although this action does not appear
naively gauge-invariant, one can use the nilpotency of the gauge group to
show that its variation under gauge transformations is a total derivative.
In the traditional reduction associated with standard (abelian) Toda theory
one gauges the maximal nilpotent algebra generated by all
the positive roots of $g$. It was then realised that one could generalise
this construction by gauging some smaller set of currents, and
moreover, that this set could be succinctly
labelled by some $su(1,1)$ embedding \cite{BTvD}.

In order to establish notation we shall see how this works in a little
more detail.
Let us consider some modified Cartan-Weyl basis for $g$,
\eq
{g=g^-\oplus h\oplus g^{+}}
\;.
\en
Here
\eq
g^\pm = \bigoplus_{\alpha\in \Delta_+} C E^{\pm\alpha} \;, \; h = \oplus C H^i
\;,
\en
where $\Delta_+$ is the set of positive roots, and
commutation relations
\eq
\comm{E^\alpha}{E^{-\alpha}} = (2/\alpha^2) \alpha^i H^i \;, \;
\comm{H^i}{E^\beta} = \beta^i E^\beta
\;.
\en
One can always conjugate any $su(1,1)$ subalgebra of $g$ so that
$I_+\in g^+$, $I_-\in g^-$ and $I_0\in h$
where
$I_+,I_-,I_0$ are
the usual raising, lowering and diagonal basis of $su(1,1)$.
We may write
$I_0=\rhov\cdot H$.
If we use the standard normalisation for the $su(1,1)$ algebra,
\eq
{}~[I_0,I_\pm]= \pm I_\pm\;,\; [I_+,I_-] = \sqrt 2I_0 \;,
\en
then
we may define the {\it characteristic} of the
$su(1,1)$ embedding to be $(\rhov\cdot e_1,...,\rhov \cdot e_i)$,
where $e_j$ are the simple roots of $g$. It is a fact that the
entries of the characteristic are $0,1/2,1$ \cite{Dynk}. We shall restrict our
attention to integral embeddings for which they must either be $0$ or $1$.
The standard reduction is associated with the principal embedding whose
characteristic contains all ones.

We may grade $g$ with respect to the $\rhov\cdd H$ eigenvalue as
\eq
g = \bigoplus_{m\in Z} g_m \,.
\en
We denote the subalgebra $\oplus_{n\ge 0}g_n$ by $p^+$ and the subalgebra
$\oplus_{n>0}g_n$ by $n^+\subset g^{+}$. We define
$p^-,n^-$ in a similar way. The above grading extends to the affine algebra in
an obvious way, and so each of the above subalgebras have `hatted'
counterparts.
We denote the restricted set of roots
$\{\alpha\in \Delta_+: E^{\alpha}\in n^+\}$ by $\delta_+$.
For the standard reduction associated with the principal reduction,
$n^\pm=g^{\pm}$ and $\delta_+=\Delta_+$.

Let us now return to the action (\ref{eq.gwzwds}). We restrict $A_L\in n^+$,
$A_R\in n^-$ and take $I_\pm$ to be the $su(1,1)$ generators as above.
Choosing the gauge $A_L=A_R=0$, the action (\ref{eq.gwzwds}) reduces to
(\ref{eq.wzw}), together with the constraints
\eq
J_L=I_++X_L(z)\;\;\;, \;\;\; J_R=I_++X_R(z)
\en
where $X_L,X_R\in {\hat p}^-,{\hat p}^+$ respectively. From now on, we shall
focus on the left chirality for ease of exposition.
Since the grading of $g$ implies that
$[g_m,g_n]\in g_{m+n}$ and $I_+\in g_1$, it follows that these constraints
are first class. Indeed they generate the gauge transformations $u\to uN(x_+)$
under which (\ref{eq.gwzwds}) is invariant, where $N\in N^-$, the group
generated by the algebra $n^-$.
Under such a gauge transformation
\eq
J_L\to J_L'={\hbar k\over 4}
(uN)^{-1}\partial_+(uN)= N^{-1}(J_L+{\hbar k\over 4}
\partial)N\, .
\label{eq.gtds}
\en
We can remove this gauge freedom by gauge fixing; this is done by further
restricting the form of $J_L$. For our purposes, the most natural gauge
is given by the {\it highest weight} gauge where we choose
\eq
J_L^{hw}=I_++\sum_iW^iE^{i}
\label{eq.hwg}
\en
where $E^{i}$ are elements of $n^-$ such that $[I_-,E^{i}]=0$
and each $E^{i}$ belongs to a distinct irreducible representation of the
grading $su(1,1)$.

It can be shown that there exists a unique solution to the equation
$J_L'=J_L^{hw}$ for $W^i$ and the gauge transformation $N$ in terms of $J_L$.
Since (\ref{eq.hwg}) specifies a well-defined point on the orbit of $J_L$
under the action of the gauge group, it follows that the components $W^i$
are gauge invariant polynomials in $J_L$. As such their Poisson brackets,
calculated using the classical version of the Kac-Moody algebra (\ref{eq.kma}),
are unaffected by the constraints and form a closed
classical W-algebra. The
number of generators of the algebra is given by the number of irreducible
representations $i$ in the decomposition of the adjoint representation of $g$
with respect to the grading $su(1,1)$. This always contains a copy of the
$su(1,1)$ itself, $\{I_+,I_0,I_-\}$, and we label this representation by $i=1$.
The coefficient $W^1$ of $E^{1}=I_-$ is the left-moving component of
the energy-momentum tensor and the other fields $W^i$ transform as primary
fields with respect to $W^1$ \cite{BFFOW1}; that is, in modes we have
\eq
\{ W^1_m,W^i_n \}_{P.B.}= [(h_i-1)m-n]W^i_{m+n}\,.
\label{eq.vprim}
\en
The conformal weight $h_i$ of the field $W^i$ is equal to $s_i+1$ where
$s_i$ is the spin of the representation $i$.
This is the beauty of the DS reduction construction of
W-algebras; the number and conformal weight of the generating fields are
given by simple Lie-algebraic considerations.

We conclude this review of the general DS reduction by making some brief
remarks about the free field representation of W-algebras
associated with generalised reduction.
It turns out that one does not need to use
all the components of $J_L$ to produce a representation of the W-algebra,
as we may have guessed, since the number of such components is greater
than the number of irreducible representations $i$. Instead, if we start
with $J_L$ in the {\it free field} gauge,
\eq
J_L^{ff}=I_++X_0
\en
where $X_0\in {\hat g}_0$ and solve $(J_L^{ff})'=J_L^{hw}$, we find that
the polynomials $W^i(J_L^{ff})$ obey the same algebra as $W^i(J_L)$.
Thus we can construct $W^i$ out of currents in ${\hat g}_0$. In the
standard reduction $g_0=h\equiv u(1)^{rank \,g}$,
the Cartan subalgebra of $g$, whence the nomenclature `free field
representation' derives. In the more general case, $g_0$ is non-abelian,
and so $W^i$ is represented using a non-abelian Kac-Moody algebra. Note
that $g_0$ is never semi-simple, as $I_0$ is always a commuting $u(1)$.
Also, the dimension of $g_0$ is equal to the number of generating fields
$W^i$, as each irreducible representation of $su(1,1)$ has one `highest
weight' component $E^{i}$ and one component of zero charge with
respect to $I_0$.

Not only can one represent the $W^i$ in terms of currents associated with
$g_0$, it is also possible using the constraints and the Polyakov-Weigmann
identity to rewrite the action (\ref{eq.gwzwds}) in
terms of a field $u_0$ taking values in the associated group $G_0$
\cite{TFeh1}.
The action is given by
\eq
S_{GT}=S(u_0)+{\hbar k\over 4}\int d^2x {\rm Tr} (I_+ u_0 I_- (u_0)^{-1})
\label{eq.gToda}
\en
For the standard reduction, the first term corresponds to the kinetic term
for rank $g$ free bosons, while the second term produces the familiar sum
of exponential terms associated with the $g$ Toda theory.
One can easily generalise the arguments given in \cite{KWat1}
to show that polynomials
in the currents constructed from $u_0$ will be chiral if and only if they
Poisson commute with the the second `potential' term. Thus this term has the
interpretation of being a generalised screening charge for the W-algebra. In
fact, we have one such screening charge for every irreducible
component of $I_-$ under the action of $g_0$. (In the standard case there are
rank $g$ such components, each belonging to one-dimensional representations
of $u(1)^{rank\, g}$.)

\section{An example of a W-algebra from generalised DS reduction}

In this section we shall introduce an example of a W-algebra which arises
from the sort of generalised reduction procedure described in the last section.
Our example has the virtue of being the simplest such algebra which still
retains most of the features of the general case.
It is therefore useful to study this algebra, both to verify existing
conjectures concerning the quantisation of generalised reductions, and to
uncover new properties which can then be generalised.

The example we shall consider arises as the reduction of a $B_2$ WZW model.
As explained above, a different model can be constructed for each
non-isomorphic
embedding of $su(1,1)$ in $B_2$. There are precisely three such, with
Dynkin indices of embedding one, two and four respectively.
The $su(1,1)$ of index
four is the principal three-dimensional subalgebra of $B_2$, associated with
the standard reduction of $B_2$. The W-algebra for this case is generated
by one field of spin two and one field of spin 4, and has already been studied
in some detail in the literature \cite{KWat2,Bajn}.
The $su(1,1)$ of index one corresponds to
the three dimensional subalgebra whose root is simply a long root of $B_2$.
The embedding is non-integral, and the corresponding algebra has generators
which do not obey the usual spin-statistics relation.

We shall concentrate on the third embedding of index two, whose root is
a short root of $B_2$.
In terms of two-dimensional Cartesian coordinates $(x_1,x_2)$, the
positive roots of $B_2$ can be taken to be
$\alpha_1=(-1,1)$, $\alpha_2=(1,0)$, $\alpha_3=\alpha_1+\alpha_2$ and
$\alpha_4=\alpha_1+2\alpha_2$. Denote the generator corresponding to
the root $\alpha_i$ by $E^{\alpha_i}$, and the Cartan subalgebra element
associated with the direction$x_i$ by $H_i$.
We identify the $su(1,1)$ generators with
$I_+=E^{\alpha_3}$, $I_-=E^{-\alpha_3}$ and $I_0=H_2$. The grading induced
by $I_0$ is just the projection onto the $x_2$ component. The
group that we are gauging is generated by the step operators associated
with the set of roots $\delta_+=\{\alpha_1,\alpha_3,\alpha_4\}$.

The $B_2$ algebra decomposes
into three spin one representations and a spin zero representation under the
action of this $su(1,1)$. By the results of the previous section it follows
that there are four generators of the W-algebra associated with this reduction:
three of conformal weight two and a single generator of weight one.
This last field forms a $u(1)$ Kac-Moody algebra, and simply corresponds
to the current associated with the $H_1$ generator which survives the
reduction.
The three spin two generators can be taken to have charges $1,0,-1$ with
respect to this global part of this $u(1)$ field, and we can identify the
chiral component of the energy-momentum tensor $L(z)$ with the field of
zero charge.

These general features of the W-algebra, together with the
requirement that the algebra be associative are sufficient to determine
its commutation relations, which we give below.
We shall call this W-algebra $\Wb$.

\subsection {The commutation relations of $\Wb$ }

In what follows we derive the commutation relations for the
new W-algebra
based on $B_2$ by
writing down
the general form of the commutation relations and checking Jacobi's identity.
The algebra contains the semi-direct product of
a $U(1)$ Kac-Moody algebra with the Virasoro algebra:
\eqq
{}~[U_m,U_n] &=& km\delta_{m+n,0} \label{eq.km}
\\ ~[L_m,L_n] &=& (m-n)L_{m+n}+{c\over 12}m(m^2-1)\delta_{m+n,0}
\label{eq.vir}
\\
{}~[L_m,U_n]&=&-nU_{m+n}
\label{eq.hw}
\enn
The algebra has two further spin-2 generators $L^+$ and $L^-$ with $U(1)$
charge $\pm 1$; that is
\eqq
{}~[L_m,L^\pm_n] &=& (m-n)L^\pm_{m+n}\\
{}~[U_m,L^\pm_n] &=& \pm L^\pm_{m+n}
\enn
By charge conservation it is easy to see that $[L^+_m,L^+_n]$ and
$[L^-_m,L^-_n]$ both vanish (since the commutator must close on fields of
spin three or less and with charge $\pm 2$, and clearly there are none),
so that the one non-trivial commutator that we need to determine is
$[L^+_m,L^-_n]$. We can use Virasoro and
Kac-Moody Ward identities to ensure that the operator product expansion
of the two generators
\eq
L^+(z)L^-(\zeta) = \sum_{n\geq 0} \psi_n(\zeta) (z-\zeta)^{-4+n},
\label{eq.wope}
\en
transform covariantly. These identities are respectively
\eqq
L_m|\psi_n\rangle &=& (m+n-2)|\psi_{n-m}\rangle
\label{eq.vw}\\
U_m|\psi_n\rangle &=& |\psi_{n-m}\rangle,
\label{eq.kmw}
\enn
for $m>0$ and the state $|\psi\rangle$ is given by the usual correspondence
$|\psi\rangle=\lim_{z\to 0}\psi(z)|vac\rangle$.
Using these relations and Jacobi's identity for the double commutator
of $L^+,L^+,L^-$ we find the following commutation relation
\eqq
{}~[L^+_m,L^-_n] &=& k^2(k-1)m(m^2-1)\delta_{m+n,0}
+k(k-1)(m^2-mn+n^2-1)U_{m+n}\cont
+ (m-n)\left [ -k(k+1)L_{m+n}+(2k-1)(U^2)_{m+n} \right ]\cont
-2(k+1) ( LU-{1\over 2}\partial^2 U )_{m+n} +2U^3_{m+n}
\enn
together with the relation
\eq
c={{-12k^2+16k-2}\over {k+1}}.
\label{eq.cch}
\en

\subsection{The free-field representation}

As mentioned in the preceding section, the classical Miura transformation
for a generalised DS reduction is of the form
\eq
N(I_++ X_0(z)+\partial_z)N^{-1} = I_++ \sum_i W^i(z)E^{i}
\label{eq.Miura}
\en
where $X_0(z)\in \hat g_0$, $[I_+,E^{i}]=0$,
and $N\in N^-$ is some gauge transformation.
In the present case, we have
\eq
\sum_i W^i(z)E^{i}=U(z)H_1+L^-(z)E^{-\alpha_1}+L(z)E^{-\alpha_3}
+L^+(z)E^{-\alpha_4}
\en
and we write $X_0=j_iH_i+j_-E^{\alpha_2}+j_+E^{-\alpha_2}$.
One can solve (\ref{eq.Miura}) for $L,L^\pm,U$ in terms of $j_\pm,j_i$.
We shall refer to the $su(2)$ subalgebra $E^{\pm \alpha_2},H_1$
associated with the
currents $\{j_\pm,j_1\}$ as the
{\it horizontal} $su(2)$, to distinguish it from the grading $su(1,1)$
whose generators are $I_\pm=E^{\pm \alpha_3},I_0=H_2$.
In the case in hand, we start by picking some matrix representation of the
algebra $B_2$, and then solve the above matrix equation explicitly.
The result is as follows, where we have ignored
coefficients, since from experience one expects these to be renormalised on
quantisation:
\eqq
U &=& j_1\\
L &=& (j_2)^2 +\partial j_2 + j_+j_-\\
L^+ &=& (j_1+j_2)j_+ + \partial j_+\\
L^- &=& (-j_1+j_2)j_- + \partial j_-
\enn
Note that we might have expected a $(j_1)^2$ term in $L$, and indeed
we shall find one below. However this ambiguity is already well understood
\cite{Dub2}.
In fact we could have guessed this answer by considering the most general
fields of correct charge and conformal weight.

In order to quantise the generalised Miura transformation given above,
we start by considering
the Wakimoto representation of $B_2$. In this approach, one constructs a
representation for the algebra $\hat g$ in terms of the currents associated
with
${\hat g}_0$ and ghosts $\beta^i,\gamma_j$ of weight one and zero
respectively, which satisfy the usual relation
\eq
\beta^i(z)\gamma_j(\zeta) \sim {\delta^i_j\over (z-\zeta)}.
\en
It is a relatively straightforward matter to write down the expressions for
$p^+=\sum_{m\ge 0} g_m$. They are given as follows:
\eqq
{\rm Tr}(JE^{-\alpha_1}) &=& \beta^1\\
{\rm Tr}(JE^{-\alpha_3}) &=& \beta^3\\
{\rm Tr}(JE^{-\alpha_4}) &=& \beta^4\\
{\rm Tr}(JE^{\alpha_2}) &=& j_+ + \beta^3\gamma_1+ \beta^4\gamma_3\\
{\rm Tr}(JE^{-\alpha_2}) &=& j_- + \beta^3\gamma_4+ \beta^1\gamma_3\\
{\rm Tr}(J H_1) &=& j_1 -\beta^1\gamma_1+\beta^4\gamma_4\\
{\rm Tr}(J H_2) &=&
\sqrt{k'+3}j_2 +\beta^1\gamma_1+\beta^3\gamma_3+\beta^4\gamma_4
\label{eq.Wakimoto}
\enn
In order for the components of the currents defined above satisfy a
Kac-Moody ${\hat B}_2$ at level $k'$, we must have that $j_\pm, j_1$
must be an $\widehat{su}(2)$ Kac-Moody algebra with central term
$k'+2$ with the length of the root squared equal to one.
This is equivalent to a $\widehat{su}(2)$ Kac-Moody algebra at level $2(k'+2)$
in the standard normalisation.The current $j_2$ is taken to be a commuting
$\hat u(1)$ with central term normalised to $k'+2$.
Ensuring that each of the above currents transforms as a spin one field
fixes the energy-momentum tensor of $B_2$ to be
\eq
T(z)=-\sum_{i=1,3,4}\beta^i\gamma_i + {\cal L}_{2(k'+2)}+{(j_2)^2\over 2}
-{3\over 2\sqrt{k'+3}}\partial j_2\,.
\label{eq.wakvir}
\en
where ${\cal L}$ is the Sugawara construction for the $su(2)$ given by
$j_\pm,j_1$. The central charge of $T(z)$ is given by
\eq
c=6+{3(k'+2)\over {k'+3}}+1-{27\over {k'+3}}={10k'\over k'+3}
\en
in agreement with what we expect for a $B_2$ Kac-Moody algebra at level $k'$.

The effect of the reduction is to remove the ghosts (quartet confinement),
and to improve the energy momentum so that the constrained currents
$\{{\rm Tr}(JE^{-\alpha_i}):i=1,3,4\}$ have weight zero.
The reduced energy-momentum tensor is
\eq
L(z)={\cal L}_{2(k'+2)}+{(j_2)^2\over 2}+\big \{\sqrt{k'+3}-
{3\over 2\sqrt{k'+3}}\big \}\partial j_2
\label{eq.impvir}
\en
The level of $U$ and $j_1$ must coincide so
we equate $k=k'+2$. It is then easy to show that the central charge of
$L(z)$ coincides with the expression (\ref{eq.cch}).

 From the classical expressions for $L^\pm$ we expect the `free field
representation of these fields to be of the form
$L^\pm=(j_1+r^\pm j_2)j_\pm + s^\pm\partial j_\pm$ where $r^\pm,s^\pm$ are
coefficients which need to be determined. Demanding that $L^\pm$
transform appropriately with respect to $L(z),U(z)$ is in fact sufficient
to determine these coefficients, and we find that
\eqq
L^+ &=& (j_1-\sqrt{k+1}j_2)j_+ -k\partial j_+\\
L^- &=& (-j_1-\sqrt{k+1}j_2)j_- -k\partial j_-\,.
\label{eq.freecharge}
\enn
These expressions together with (\ref{eq.impvir}) and $U=j_1$ satisfy the
commutations of $\Wb$ given above, at least up to normalisation.

Often the free field realisation can also be constructed as the commutant
of a set of screening charges \cite{FLuk}.
In the previous section we saw that the
classical expression for the screening charge is given by
$\int d^2x {\rm Tr} (I_- u_0 I_+ (u_0)^{-1})$.
Since $I_+=E^{\alpha_3}$ transforms as a spin one field under the action
of the horizontal $su(2)$, we expect the screening charge(s) to be
the integral of a primary field(s) transforming as a $(3,-)$ under
$g_0=su(2)\times u(1)$,
with zero horizontal charge and with conformal weight one.
If $\psi^{adj}(z)$ is the zero charge component of
a primary field for the horizontal $su(2)$ Kac-Moody
algebra transforming in the adjoint representation,
then the quantum screening charge $S$ is given by
\eq
S=\int dz \psi^{adj}(z)exp[\alpha iX_2(z)]
\label{eq.screening}
\en
where $j_2(z)=i\partial X_2(z)$. One determines $\alpha$ from the
condition that $S$ must have conformal weight one by using equation
(\ref{eq.impvir}). We find the two solutions
\eq
\alpha=-{1\over \sqrt{k+1}}\, ,\, {2k\over \sqrt{k+1}}\, \, .
\en
It is then a relatively straightforward if tedious calculation to show
that the screening charges $S$ commutes with the free field expressions
for the generators of $\Wb$ given above if we choose the first of the
two solutions for $\alpha$ and we relegate it to an appendix.

\section{Basics of $\Wb$ representation theory}

In order to set up the representation theory for $\Wb$, we first have to
understand what a `highest weight' representation for this algebra is.
Loosely speaking, a highest weight representation is one such that the
representation space $V$ can be graded with respect to $L_0$ eigenvalue
\eq
V=\bigoplus_{m\ge 0}V_m
\en
where the space $V_m$ has $L_0$ eigenvalue $h+m$. Here we concentrate on
the space of highest weight states $\psi\in V_0$. Since $V_m=\emptyset$
for $m<0$, it follows that
\eq
X_m|\psi\rangle=0\,\, ,\,\, m>0
\label{eq.hw}
\en
for any generator $X_m$ of $\Wb$.

Now let us consider the action of the zero modes $X_0$ on $V_0$. Algebraically,
the zero modes of $\Wb$ do not close; for instance
\eq
{}[L^+_0, L^-_0] = ... - 2(k+1)(LU)_0
=...-2(k+1)[\sum_{x>0} U_{2-x} L_{x-2}+\sum_{x>0} L_{-1-x} U_{x+1}].
\en
However, on $V_0$, the non-zero modes of $\Wb$ vanish because of (\ref{eq.hw}),
and indeed the algebra restricted to the zero modes is a consistent,
associative
algebra. The mode $L_0$ commutes with $X_0$, and so simply acts as a central
term with value $h$ in $V_0$. The remaining three zero modes of generating
fields of $\Wb$ have the following  commutation relations;
\eqq
{}[U_0,L^\pm_0] &=& \pm L^\pm_0 \\
{}[L^+_0,L^-_0] &=& \big [ -k(k-1)-2(k+1)h \big ] U_0 + 2(U_0)^3\, .
\label{eq.dsu2}
\enn
Thus we see that these three modes form a sort of deformation of the
usual $su(2)$. We call this algebra $\Wb_0$.
This is very like the quantum group $su_q(2)$, but in
this case the commutators close on polynomial terms, rather than on
hyperbolic functions. Polynomial algebras of this sort
have been considered before \cite{Tjin1}, where they were called {\it finite
W-algebras}. In fact the algebra (\ref{eq.dsu2}) has a longer history
\cite{Zhed,Higg}
appearing as the algebra of conserved quantities for a Coulombic central
force problem on a space of constant curvature.

In analogy with the representation theory of $su(2)$, we shall consider
representations built up from a highest weight state $\psi_\Lambda$,
with the properties that
\eqq
L^+_0|\psi_\Lambda\rangle &=& 0\\
U_0|\psi_\Lambda\rangle &=& \Lambda |\psi_\Lambda\rangle
\label{eq.suhw}
\enn
Thus we specify representations of $\Wb$ by three parameters: $c$
(or equivalently $k$), the weight $h$ of the highest weight
state $\psi_\Lambda$, and its charge $\Lambda$.

The space $V_0$ is built up by applying the `raising operator' $L^-_0$ to
$\psi_\Lambda$, and
the states $(L^-_0)^p|\psi_\Lambda\rangle$ form a basis for $V_0$.
For arbitrary $\Lambda$, the space $V_0$ is therefore an infinite-dimensional
representation of $\Wb_0$. For special $\Lambda$ though, we can arrange that
the norm of $(L^-_0)^p|\psi_\Lambda\rangle$ vanishes. To see when this occurs,
note that
\eq
|(L^-_0)^{p+1}|\psi_\Lambda\rangle|^2
=
\sum_{i=0}^{p}P(\Lambda-i,h)
|(L^-_0)^{p}|\psi_\Lambda\rangle|^2
\label{eq.znorm}
\en
where $P(\lambda,h)$ is simply the right hand side of (\ref{eq.dsu2}) evaluated
on a state of charge $\lambda$ and weight $h$; i.e.
\eq
P(\lambda,h)=[-k(k-1)-2(k+1)h]\lambda+2\lambda^3\, .
\en
Since $P$ is odd under $\lambda\to -\lambda$, it follows that the norm vanishes
for $\Lambda\in {{\bf Z}\over 2}$ and $p=2\Lambda+1$, just as for standard
$su(2)$. There are other solutions though. To express these in a compact form,
it turns out it is useful to use a parametrisation of $h$ inspired by the
free field representation we found in the last section. Any highest weight
state $\psi_{ff}$ for ${\hat g}_0$ is obviously
also a highest weight state for $\Wb$. If
\eqq
(j_1)_0|\psi_{ff}\rangle &=& {\Lambda}|\psi_{ff}\rangle\\
(j_2)_0|\psi_{ff}\rangle &=& {a\over \sqrt{k+1}}|\psi_{ff}\rangle
\enn
then $\psi_{ff}$ is a highest weight state for $\Wb$ with charge $\Lambda$ and
\eq
h={{\Lambda(\Lambda+1)+a(a+1-2k)}\over 2(k+1)}
\label{eq.hff}
\en
where we have used (\ref{eq.impvir}) to find this expression. From now on,
we shall use $a$ rather than $h$ to parametrise the conformal weight
of the highest weight state. We then find that the norm of descendant of
the state $|\psi_{ff}\rangle=|\Lambda,a\rangle$ is
\eqq
|(L^-_0)^p|\Lambda ,a \rangle|^2&=&N^{p-1}|(L^-_0)^{p-1}|\Lambda,a\rangle|^2\\
N^{p-1}&=& -{p\over 2}(2\Lambda+1-p)
(p-\Lambda-k+a)(\Lambda+1-p-k+a)
\label{eq.normzero}
\enn
so that we can expect $V_0$ to be finite-dimensional if and only if at least
one of $2\Lambda+1$, $\Lambda+k-a$ or $\Lambda+1-k+a$ is a positive integer.

We conclude our discussion of zero modes by mentioning that it is possible
to define a Casimir operator $C$ for $\Wb_0$, given by \cite{Zhed}
\eq
C =
L^-_0 L^+_0 - {1\over 2}[k(k-1)+2(k+1)h] U_0 (U_0+1)+{1\over 2} U_0^2
(U_0+1)^2
\label{eq.Casimir}
\en
In particular we find using the parametrisation (\ref{eq.hff}) that
\eq
C|\Lambda,a\rangle = -{1\over 2}
\Lambda(\Lambda+1)(a-k)(a-k+1)|\Lambda,a\rangle\, .
\label{eq.cff}
\en

The pair $(C,h)$ are an invariant way of specifying an irreducible
representation
of $\Wb$. Let us compare this with the Casimirs of the other two reductions of
$B_2$ with which we are familiar: the `null' reduction or just the Kac-Moody
algebra ${\widehat{so}}(5)$ and the standard reduction with algebra $WB_2$.
All three algebras possess two Casimirs which are polynomials of order two
and four in terms of the free field parametrisation ${\lambda}$
of their representations \cite{KWat2}.
Furthermore these polynomials exhibit a symmetry under a shifted
$B_2$ Weyl group action ${{\lambda}}-{{\rho}}\to
w({{\lambda}}-{{\rho}})$ for some constant vector $\rho$. For
$\Wb$ $\rho=(-1/2,-1/2+k)$ and setting $(\Lambda',a')=(\Lambda+1/2,a+1/2-k)$
we see that $C$ and $h$ as given in (\ref{eq.hff}), (\ref{eq.cff}) are
indeed invariant under the $B_2$ Weyl group action
\eq
(\Lambda',a')\to (\pm \Lambda', \pm a')\, \, .\, \, (\pm a', \pm \Lambda')
\label{eq.wga}
\en
Note that if we express the formula for the norm of $(L^-_0)^p|\Lambda,a
\rangle$ (\ref{eq.znorm}) in terms of its charge $q=\Lambda-p$ then it
is also invariant under the action of the shifted Weyl group.

\section{Representation theory of $\Wb$ from fusion}

In the last section we saw how to construct highest weight representations
of $\Wb$ as Verma modules built up from a highest weight state $|\Lambda,a
\rangle$. For generic values of $\Lambda$ and $a$ these representations
will be irreducible. However, past experience teaches us that the physically
interesting cases are the special cases where the Verma module is not an
irreducible representation of $\Wb$ but instead contains one or more singular
vectors. Such representations are called {\it degenerate}. Indeed, we have
already entertained the possibility that $|\Lambda,a\rangle$ has a singular
vector at level zero (\ref{eq.znorm}), since this avoids the rather
unpleasant infinite degeneracy in energy that would arise otherwise.

In general we expect that each independent singular vector (i.e. a singular
vector which is not a descendent of other singular vectors)
in the Verma module requires
an equation to be satisfied by the parameters specifying the representation.
Therefore the maximum number of such independent singular vectors is equal to
the number of parameters specifying the representation, and in analogy with
Lie algebras of finite dimension we refer to this as the {\it rank} of the
W-algebra. Although we could (and perhaps should) include the central charge
$c$ (or in our case the value of $k$) in the parametrisation, in this paper
we choose to exclude it from the set and leave it as an arbitrary irrational
number, since the representation theory is simplest to work out in this
`quasi-rational' case.
We expect that for rational $c$, representations can have many more singular
vectors, and that the fusion rules acquire a quantum-group like structure.
We call representations
containing the maximum number of independent and dependent singular vectors
{\it maximally degenerate} representations (although not all
representations with the maximum number of independent null vectors are
maximally degenerate). We reserve the nomenclature {\it completely degenerate}
for the equivalent concept for rational $c$.
 From previous experience maximally degenerate representations should have
the special property that the fusions of
the corresponding primary fields can be completely determined by demanding
that their null descendents decouple and so we shall be primarily interested
in such representations.

The representations of $\Wb$ are specified by the pair $(\Lambda,a)$ or
equivalently by the Casimirs $(C,h)$ (where as mentioned above we take
$k$ or $c$ to be irrational so that this parameter does not play a role in
the singular vector structure). Thus the rank of $\Wb$ is two, and
so maximally degenerate representations have two independent singular vectors.
If we demand that the highest weight space $V_0$ form a finite dimensional
representation of $\Wb$, then we must have an independent  singular
vector in $V_0$, leaving
in general one more independent singular vector in $V_m$, $m>0$.
In the remainder of this section we shall investigate the fusion of
various basic representations which are of this type.

\subsection{Using null vectors to determine fusion}

We follow the approach developed in \cite{BDIZ1,BPTa1,BWat3}.
This involves choosing representations
whose independent singular vector is in $V_m$ for m small.
This vector is explicitly constructed
by brute force and then used to solve for fusion of the
corresponding primary field with any other primary field. As a side product,
we also find an infinite set of representations which are degenerate,
and find a recurrence relation which can be used to
explicitly construct the singular vector associated with these representations.

We begin by considering the classical form of null vectors, for the simple
case of the $B_2$ WZW model. Let us recall the form of the currents
(\ref{eq.currents}). This can be re-written
\eq
(\partial_z - J^a_LT^a)g=0\, \, a\in g
\label{eq.kz}
\en
where we take $T^a$ to be some particular representation $\lambda$ of $g$.
Although $g$ is a matrix, the equation (\ref{eq.kz})
acts on each column of $g$ independently, so we can think of $g$ as
being a column vector (we ignore the other chirality for the moment).
 From now on we write $g=g_j$ where $j$ runs over the different elements
of the representation $\lambda$.
The trick is to think of (\ref{eq.kz}) as the classical version of
the Kniznik-Zamolodchikov equation
\eq
L_{-1}- J^a_{-1}T^a|\psi\rangle
\label{eq.qkz}
\en
where $\psi$ is some highest weight state for ${\hat g}$.
Here and below we ignore constant coefficients for simplicity.
This is a singular vector which is in any highest weight representation
of the semi-direct product of a Kac-Moody and Virasoro algebra.

When we reduce these equations, after gauge fixing the equation
(\ref{eq.kz}) becomes
\eq
(\partial_z\delta_{jk} - (I_+)_{jk}-\sum_i W^i(E^{i})_{jk})g_k=0 \, .
\label{eq.classicaleqm}
\en
This set of coupled differential equations can be thought of as
the classical equivalent of some null vector condition for a
representation of the W-algebra. The order of the system of
equations is simply equal to the dimension of the representation that
$g$ is in, and one can think of different fusions of $g$ as being
associated with different solutions of this equation. It is therefore
very clear at the classical level that there is a close correspondence
between representations of finite Lie algebras and the representations
of W-algebras obtained by DS reduction.

The quantised version of (\ref{eq.classicaleqm}) is the set of equations
\eq
(L_{-1}\delta_{jk} - (I_+)_{jk}-\sum_i W^i_{-h_i}
(E^{i})_{jk})|\psi_k\rangle=0 \, ,
\label{eq.quantumeqm}
\en
where we have ignored some constant coefficients which experience shows need to
be introduced. This equation simplifies if we choose a basis for the
representation $\lambda$ which diagonalises the action of the Cartan
subalgebra $H^i\in h$, so that we have
\eq
H^i|\psi_k\rangle=\lambda_k^i|\psi_k\rangle
\en
We can extend the grading of the Lie algebra $g$ introduced in section
two to the states $\vec {\psi_k}$ by defining
\eqq
G(H^i) &=& 0\\
G(E^{\alpha_i}) &=& \rhov \cdot \alpha_i \\
G(\vec {\psi_k}) &=& \rhov \cdot \lambda_k\, \, .
\enn
Consistency of (\ref{eq.quantumeqm}) with respect to $L_0$ shows that
\eq
L_0|\psi_k\rangle=(\kappa-G(|\psi_k\rangle))|\psi_k\rangle
=(\kappa-\lambda_k\cdot \rho)
|\psi_k\rangle.
\en
where $\kappa$ is a constant. We take $|\psi_k\rangle
\in V_m$ if $(\lambda-\lambda_k)\cdot \rho=m$. In particular if $|\psi_{
\lambda}\rangle$ is the state corresponding to the highest weight $\lambda$
then $|\psi_{\lambda}\rangle \in V_0$ and is also a highest weight state
for $g_0$,
so we identify this as the highest weight state for the W-algebra.

The null vectors implied by (\ref{eq.quantumeqm}) can be used
to solve explicitly for
the operator product expansion of $|\psi_\lambda\rangle$ with some other
primary field $\phi$. Corresponding to each null state $|\chi\rangle$ is
a null field $\chi(z)$ and we simply demand that the operator product
expansion of this field with $\phi$ is zero or a
sum of null fields. Using the methods of \cite{BDIZ1,BPTa1,BWat3},
we can reexpress this condition
as an equation in terms of the operator product of $\psi_\lambda$ with $\phi$,
and this enables us to solve for the latter. As an example, consider
the operator product of the descendant field corresponding to
$(L^+)_{-p}|\psi_\lambda\rangle$ with some other primary field $\phi$. This
can be written
\eqq
\hat L^+_{-p}\psi_\lambda(z) \vec \phi
&=& \oint_z d\zeta  L^+(\zeta)\psi_\lambda(z)(\zeta-z)^{1-p}\vec \phi\\
&=& \int_{\zeta>z}-\int_{\zeta<z} d\zeta  L^+(\zeta)
\psi_\lambda(z)(\zeta-z)^{1-p}\vec \phi\\
&=& \left[
\sum_{s=0}^\infty  \pmatrix{ p-2-s \cr s } z^s L^+_{-p-s}
\psi_\lambda(z)+(-1)^{p} z^{1-p-s} \psi_\lambda(z) L^+_{-1+s}
\right]
\vec \phi\\
&=& \left[
\sum_{s=0}^\infty  \pmatrix{ p-2-s \cr s } z^s L^+_{-p-s}
- (-1/z)^{p-1}(   L^+_{-1} - \hat L^+_{-1})
\right]
\psi_\lambda(z)\vec \phi
\label{eq.bsa}
\enn
By similar manipulations it is possible to rewrite the operator product
expansion of any descendent field ${\hat X}_{-m}\psi_{\lambda}|\phi\rangle$
as a sum of terms involving operators acting on terms of the form
\eq
O(i,p,q,r;z)=({\hat L}^+_{-1})^p({\hat L}^-_{-1})^q\psi_{\lambda}
({\hat L}^-_0)^r
|\phi\rangle,
\label{eq.opebasis}
\en
where $p+q\le m$ and $p-q-r$ equals the charge of the operator ${\hat
X}_{-m}$.
The $O(p,q,r;z)$ can be thought of as forming a basis of
independent operator product expansions. For maximally degenerate
representations, we conjecture that there are precisely enough null vectors
in the Verma module to solve for the $O(i,p,q,r;z)$.

We shall now apply this technique to the two fundamental representations
of $B_2$, that is the spinor and vector representations, and solve for
the fusions of the primary fields corresponding to these representations.
In addition, we shall also find the weights of a subset of all degenerate
representations, and an explicit expression for their null vectors.

\subsection{The spinor representation of $B_2$}

The basic representation of $B_2$ is the spinor
representation, with the four weights $(\pm 1/2, \pm 1/2)$. It is easiest
to work in the eigenvalue basis of this representation, so
again ignoring constants since we expect these to be renormalised on
quantisation anyway, for this case (\ref{eq.classicaleqm}) reads
\eq
\pmatrix{U & 0 & 1 & 0\cr
        0 & -U & 0 & 1 \cr
        L & L^+ & U & 0 \cr
        L^- & L & 0 & -U}
\pmatrix{g_1 \cr g_2 \cr g_3 \cr g_4}
=
\partial
\pmatrix{g_1 \cr g_2 \cr g_3 \cr g_4}
\en
The first and second equation can be used to eliminate $g_3$ and $g_4$
in favour of $g_1$ and $g_2$. Substituting into the last two equations
we find the following coupled differential equations:
\eqq
{}[(\partial +U)^2-L]g_2 &=& L^- g_1 \\
{}[(\partial -U)^2-L]g_1 &=& L^+ g_2 \, .
\label{eq.spineqm}
\enn
The components $g_1$, $g_2$ have $U_0$ charge $1/2$ and $-1/2$ respectively.
We label the corresponding quantum states $|\psi_+\rangle$ and
$|\psi_-\rangle$.
As above, we take $\vec {\psi_+}$ to be a highest weight state for $\Wb$ and
$\vec {\psi_-} = L^-_0 \vec {\psi_+}$.
We label the conformal weight of these states $\delta_s$ so that
$L_0|\psi_\pm\rangle=\delta_s|\psi_\pm\rangle$.
 From the classical constraint equation (\ref{eq.spineqm}), we expect
there to be a null state of the form
\eq
|\chi_+\rangle=
(L^+)_{-2}|\psi_-\rangle+(\b_1L_{-2}+\b_{2}L_{-1}L_{-1}+\b_3 U_{-1}L_{-1}
+\b_4 U_{-1}U_{-1}+ \b_5 U_{-2})|\psi_+\rangle\, .
\label{eq.null2}
\en
Applying $L_1$ to this vector we see that we also expect a null vector of
the form
\eq
(L^+)_{-1}|\psi_-\rangle + (\gamma_1 U_{-1} + \gamma_2 L_{-1})|\psi_+
\rangle=0\, .
\label{eq.null1}
\en
There are two solutions to this being a null vector: $(\gamma_1,\gamma_2,
\delta_s)= (k-1,1-k^2,{{3-2k}\over {4(k+1)}})$ or
$({2k+1\over 8}, {{(k+1)(2k+1)}\over 4},{{8k+3}\over {8(k+1)}})$. The
second solution is a highest weight state for $\Wb_0$, or in other
words is annihilated by $L^+_0$. The first solution belongs to a null
spin-$3/2$ multiplet of $\Wb_0$, and is a  descendent of the
state $L^+_{-1}|\psi_+\rangle$ which is also null in this case. In fact,
only the first solution has a null descendent of the form (\ref{eq.null2}).
Explicitly this is
\eq
\b_1={k^2-1\over 2}\, , \, \b_2= (k+1)^2(1-k) \, , \,
\b_3=k^2-1\, , \, \b_4= {1-k\over 2}\, , \, \b_5={(k+2)(k-1)\over 2}
\label{eq.null2coeff}.
\en
There is also a null vector $\chi_-$ of spin $-1/2$
which is the charge conjugate of $\chi_+$.

Let us write
\eqq
\psi_+(z)(L^-_0)^p|\phi\rangle&=&\sum_{n\in {\bf Z}_+} \mu^p_n z^{n-y}\\
\psi_-(z)(L^-_0)^p|\phi\rangle&=&\sum_{n\in {\bf Z}_+} \nu^p_n z^{n-y}\,\, .
\enn
We now use the null vectors (\ref{eq.null2}), (\ref{eq.null1}) together
with the relations at level $0$,
\eqq
L^+_0|\psi_+\rangle &=& 0,\\
L^-_0|\psi_+\rangle &=& |\psi_-\rangle,\\
L^-_0|\psi_-\rangle &=& 0,
\enn
to solve for the above operator product expansions. Performing the
sort of contour manipulations described in the previous section we arrive
at the following recurrence relations:
\eqq
\alpha^p_n\mu^p_n &=&
{2N^{p-1}\over (k-1)} \nu^{p-1}_n
+\sum_{x>0}U_{-x}\mu^p_{n-x}[2(N-Y-k-1)-k(x-1)-2(\Lambda-p)]
\cont
+ {2\over (k-1)}\sum_{x>0}
L^+_{-x}\nu^p_{n-x}+\sum_{x>0}L_{-x}\mu^p_{n-x}(1+k)
-\sum_{x,y>0}U_{-x}U_{-y}\mu^p_{n-x-y}
\label{eq.spinrr}
\\
\beta^{p-1}_n\nu^p_n &=&
(1-k) \sum_{x\ge 0}
L^+_{-x}\mu^p_{n-x}+\sum_{x>0}L_{-x}\nu^p_{n-x}(1+k)
-\sum_{x,y>0}U_{-x}U_{-y}\nu^p_{n-x-y}
\cont
+\sum_{x>0}U_{-x}\nu^p_{n-x}[2(N-Y-k-1)-k(x-1)-2(p-1-\Lambda)]
\label{eq.spinrr''}
\enn
where
\eqq
\alpha^p_n &=& 2(k+1)^2(y-n)^2+(2k-1+2(\Lambda-p))(k+1)(y-n)+(\Lambda-p)
(\Lambda-p+k)\cont -(1+k)h_\phi\\
\beta^p_n &=& 2(k+1)^2(y-n)^2+(2k-1-2(\Lambda-p))(k+1)(y-n)+(\Lambda-p)
(\Lambda-p-k)\cont -(1+k)h_\phi
\enn
and $N=n(k+1)$, $Y=y(k+1)$.
Alternatively, the equation (\ref{eq.spinrr''}) can be
expressed as
\eqq
\beta^{p}_n\nu^p_n &=& (1-k)\mu^{p+1}_n+
(1-k) \sum_{x>0}
L^+_{-x}\mu^p_{n-x}+\sum_{x>0}L_{-x}\nu^p_{n-x}(1+k)
-\sum_{x,y>0}U_{-x}U_{-y}\nu^p_{n-x-y}\, .
\cont
+\sum_{x>0}U_{-x}\nu^p_{n-x}[2(N-Y-k-1)-k(x-1)-2(p-\Lambda)]
\label{eq.spinrr'}
\enn

Consistency of (\ref{eq.spinrr'}) with (\ref{eq.spinrr}) for $n=0$ implies that
$2N^{p-1}=-\alpha^p_0\beta^{p-1}_0$. Substituting in for $N^p,\alpha^p_0$ and
$\beta^p_0$, we see that this equation is equivalent to
\eqq
0=\alpha^0_0\beta^{-1}_0&=&
(2Y^2+(2k-1+2\Lambda)Y+\Lambda
(\Lambda+k) -(1+k)h_\phi)\times \cont
(2Y^2+(2k-3-2\Lambda)Y+(\Lambda+1)
(\Lambda+1-k) -(1+k)h_\phi)
\label{eq.fuseq}
\enn
With some more work we can use this equation to solve for the allowed
fusions $\psi^\pm \times \phi \to \phi'$. Taking the free field
parametrisation for $\psi^+$, $\phi$, $\phi'$
to be $(1/2,1/2)$, $(\Lambda,a)$, $(\Lambda',a')$
respectively we find that
\eq
(1/2,1/2)\times (\Lambda,a) \to (\Lambda',a')=(\Lambda\pm {1\over 2},
a\pm {1\over 2})
\label{eq.spinfusion}
\en
The fusion rule for the field $\psi^+$ is exactly the same that
selection rules for the finite dimensional spinor representation from
which it is derived. We shall see that this property holds for more
general representations below.

We can extract more information from (\ref{eq.spinrr}).
For any of the above allowed fusions (\ref{eq.spinfusion}), we can
substitute in (\ref{eq.spinrr}) the corresponding solution for $y$,
and solve explicitly for $\mu^p_n$, $\nu^p_n$ in terms of the
highest weight state ($\mu^0_0$ or $\nu^0_0$ depending on whether
$\Lambda'=\Lambda+1/2$ or $\Lambda-1/2$). However, for special values
of $(\Lambda,a)$ this iterative process breaks down because one of the
coefficients $\alpha^p_n,\beta^p_n$ vanishes. The right hand side
of the corresponding equation (\ref{eq.spinrr}) can be interpreted as a
null vector. For instance, consider
the fusion $(1/2,1/2)\times (\Lambda,a)\to (\Lambda+1/2,a+1/2)$.
In this case $y=-(\Lambda+a)/2$ and
\eqq
\alpha^p_n&=&N(2N+2a-2k+1+2p)+p(p-\Lambda-k-a)\\
\beta^p_n&=&N(2N+2a-2k+1+4\Lambda-2p)+(2\Lambda-p)(\Lambda-p-k-a).
\enn
Note that $\alpha^0_0=0$, as required by (\ref{eq.fuseq}). If
in addition $a=k-1/2-n(k+1)$, then $\alpha^0_n$ also vanishes, indicating
the presence of a null vector at level $n$ with $U_0$ charge $\Lambda+1/2$
in the module labelled by $(\Lambda',a')=(\Lambda+1/2,k-n(k+1))$. Similarly,
if $a'=k-1/2-n(k+1)-\Lambda$, then $\beta^{-1}_n$ vanishes, indicating
that there is a null vector at level $n$ with $U_0$ charge $\Lambda-1/2$.
In both cases, we can solve the recurrence relations
(\ref{eq.spinrr},\ref{eq.spinrr''})
up to level $n$, and use the level $n$ equation to find an explicit expression
for the null vector, much as was done in \cite{BWat3}.

One might be tempted to argue that the vanishing $\alpha^p_n,\beta^p_n$
for $p>0$ gives null vectors with lower $U_0$ charge, but it turns out
that the equations (\ref{eq.spinrr}), (\ref{eq.spinrr''}) conspire
to give zeroes on the right hand side which can be cancelled with those on the
left in this case.

\subsection{The vector representation of $B_2$}

The vector representation of $B_2$ is five-dimensional with weights
$(0,1)$, $(-1,0)$, $(0,0)$, $(1,0)$ and $(0,-1)$, to which we associate
an eigenvalue basis $g_1$, $g_2$, $g_3$, $g_4$ and $g_5$ respectively. In
this basis, the classical equation of motion (\ref{eq.classicaleqm}) reads
\eq
\pmatrix{0 & 0 & 1 & 0 & 0\cr
        L^- & U & 0 & 0 & 0\cr
        L & 0 & 0 & 0 & 1\cr
        L^+ & 0 & 0 & U & 0\cr
        0 & L^+ & L & L^- & 0}
\pmatrix{g_1 \cr g_2 \cr g_3 \cr g_4 \cr g_5}
=
\partial
\pmatrix{g_1 \cr g_2 \cr g_3 \cr g_4 \cr g_5}
\en
 From general considerations we identify the state $\vec {\psi_1}$ associated
with $g_1$ as a highest weight state for $\Wb$, which has zero $U_0$ charge
and therefore is a singlet under the action of $\Wb_0$.
The first and third of these equations imply that
$\vec {\psi_3}\sim L_{-1}\vec {\psi_1}$, $\vec {\psi_5}\sim (L_{-2}+\zeta
L_{-1}
L_{-1})\vec {\psi_1}$, while the weight and charge of $\vec {\psi_2}$ and
$\vec {\psi_4}$ imply that $\vec {\psi_2}=L^-_{-1}\vec {\psi_1}$
and $\vec {\psi_4}=L^+_{-1}\vec {\psi_1}$. The remaining three equations imply
the existence of three null vectors of the following form
\eqq
\vec {n^+} &=&
(\gamma_1 L^+_{-2} + \gamma_2 U_{-1}L^+_{-1} + \gamma_3 L_{-1} L^+_{-1})
\vec {\psi_1}\\
\vec {n^-} &=&
(\gamma_1 L^-_{-2} - \gamma_2 U_{-1}L^-_{-1} + \gamma_3 L_{-1} L^-_{-1})
\vec {\psi_1}\\
\vec {n^f} &=&
( L^-_{-2}L^+_{-1}+L^+_{-2}L^-_{-1} + \beta_1 L_{-1}U_{-1}U_{-1} + \beta_2
U_{-1}U_{-2} + \beta_3 L_{-3} +\beta_4 L_{-1} L_{-2}\cont
+\beta_5 (L_{-1})^3)
\vec {\psi_1}
\enn
Note that we are allowed to add terms of the form $U^2$ to the last null
vector, much as the classical and quantum energy momentum tensor by terms
of this way.

As discussed above, maximally degenerate representations with
a finite-dimensional highest weight space $V_0$ have only one independent
null vector at higher level; therefore we expect that $\vec {n^-} \sim
(L^-_0)^2\vec {n^+}$. We label the neutral null vector $L^-_0\vec {n^+}$
by $\vec {n^0}$. The charge conjugation invariance of $\vec {n^f}$ implies
that
\eq
\vec {n^f}\sim \epsilon_1 (L^+_{-1}\vec {n^-} + L^-_{-1}\vec {n^+})+
\epsilon_2L_{-1}\vec {n^0}
\en

We find two possible solutions for the above null vectors which we consider
in turn below.
\vskip 5mm

\noindent {\bf Case (i): $L_0\vec {\psi_1} = {1-k \over 1+k}\vec {\psi_1}$}

This case closely parallels the results for the spinor field in the
previous section. We find that
\eqq
\vec {n^+} &=&
(L^+_{-2} + U_{-1}L^+_{-1} - (1+k) L_{-1} L^+_{-1})
\vec {\psi_1}\\
\vec {n^0} &=& (L^+_{-1}L^-_{-1}+L^-_{-1}L^+_{-1}-4U_{-1}U_{-1}+
4(k+1)L_{-2}-2(k+1)^2(L_{-1})^2)
\vec {\psi_1}\\
\vec {n^-} &=&
(L^-_{-2} - U_{-1}L^-_{-1} -(1+k) L_{-1} L^-_{-1})
\vec {\psi_1}\\
\vec {n^f} &=&
( L^-_{-2}L^+_{-1}+L^+_{-2}L^-_{-1} -(k+1)  L_{-1}U_{-1}U_{-1} +(k-3)
U_{-1}U_{-2} - k(k+1) L_{-3}\cont
+2(k+1)^2  L_{-1} L_{-2} -(k+1)^3 (L_{-1})^3)
\vec {\psi_1}
\enn
Much as in the spinor case, where the null vectors $\chi_-,
\chi_+$ were part of a larger multiplet of null vectors with `highest
weight' $L_{-1}^+|\psi_+\rangle$, in this case
$\vec{n^+},\vec{n^0},\vec{n^-}$ form part of a spin-$2$ multiplet with
highest weight $(L^+_{-1})^2\vec{\psi_1}$.
We let
\eqq
\psi_1(z) (L^-_0)^p\vec{\phi} &=& \psi^p_n z^{n-y}\\
\hat L^+_{-1}\psi_1(z) (L^-_0)^p\vec{\phi} &=& (\phi^+)^p_n z^{n-y-1}\\
\hat L^-_{-1}\psi_1(z) (L^-_0)^p\vec{\phi} &=& (\phi^-)^p_n z^{n-y-1}
\enn
Since $L^\pm_0\vec{\psi_1}=0$ we have that
\eqq
(\phi^+)^p_n &=& L^+_0 \psi^p_n - N^{p-1} \psi^{p-1}_n\\
(\phi^-)^p_n &=& L^-_0 \psi^p_n -  \psi^{p+1}_n\, .
\label{eq.vectorrr0}
\enn
Then we can use the vectors $\vec {n^+}$, $\vec{n^-}$ to derive the
following relations:
\eqq
\alpha^p_n (\phi^+)^p_n &=& N^{p-1}\psi^{p-1}_n+\sum_{x>0}L^+_{-x}\psi^p_{n-x}
+U_{-x}(\phi^+)^p_{n-x}\\
\beta^p_n (\phi^-)^p_n &=& \psi^{p+1}_n+\sum_{x>0}L^-_{-x}\psi^p_{n-x}
-U_{-x}(\phi^-)^p_{n-x}\, ,
\label{eq.vectorrr1}
\enn
where
\eqq
\alpha^p_n &=& N-Y-k-\Lambda+p\\
\beta^p_n &=& N-Y-k+\Lambda-p\, .
\enn
We can also write (\ref{eq.vectorrr1}) alternatively
using that $L^-_0\vec{\psi_1} =0$ as
\eq
\beta^{p-1}_n (\phi^-)^p_n = \sum_{x\ge 0}L^-_{-x}\psi^p_{n-x}
-\sum_{x>0}U_{-x}(\phi^-)^p_{n-x}\, .
\en
Combining the information in $\vec{n^f}$ and $\vec{n^0}$ gives the following
equation:
\eqq
\gamma^p_n \psi^p_n &=& L^-_0 (\phi^+)^p_n + N^{p-1}(\phi^-)^p_n\cont
\sum_{x>0} (k+1)[2(N-Y)-kx-2(k-1)]L_{-x}\psi^p_{n-x}\cont
+ [-3(N-Y)+2x(k-2)-(k-1)]U_{-x}\psi^p_{n-x}\cont
\sum_{x,w>0} [-(N-Y)+(k-3)w+(k-1)]U_{-x}U_{-w}\psi^p_{n-x-w}
\label{eq.vectorrr2}
\enn
where
\eq
\gamma^p_n=\{(N-Y)-k+1\}
           \{(N-Y)^2+(1-2k)(N-Y)+(\Lambda-p)^2+(\Lambda-p)
           -2(k+1)h_\phi\}
\en

Using (\ref{eq.vectorrr2}) for $n=0$ together with equations derived
from null vectors at level zero, and expressing $h_\phi$ in terms of
$a$ and $\Lambda$ we find that
\eq
(Y+k-1)[(Y+k-1-\Lambda+p)(Y+k+\Lambda-p)(Y+a)(Y-a-1+2k)-N^{p-1}]\psi^p_0=0
\en
The $p$ dependence cancels in these equations and since by definition
$\psi^p_0 \ne 0$ for some $p$, it follows that $Y$ must satisfy
\eq
(Y+k-1)(Y+k-1-\Lambda)(Y+k+\Lambda)(Y+a)(Y-a-1+2k)=0
\en
One can also evaluate the fourth order Casimir $C$ on the states $\psi^p_0$
and show that only the following fusions are allowed:
\eq
{\psi_1}(z) \times (\Lambda,a)\to(\Lambda',a')=
(\Lambda \pm 1,a),(\Lambda,a),(\Lambda,a\pm 1)
\en
Just as in the spinor case, the allowed fusions for the degenerate
primary field derived from the vector representation correspond to the
tensor decomposition rules of the vector representation. Note that
though $\psi_1(z)$ is a singlet under $\Wb_0$ its fusion with other
primary fields changes the value of $\Lambda$. This stems from the simple
observation that
\eq
[L_0^+,\psi_1(z)]=z{\hat L}^+_0 \psi_1(z)+z{\hat L}^+_{-1}
\psi_1(z)=z{\hat L}^+_{-1} \psi_1(z)\neq 0
\en

\vskip 5mm
\noindent{\bf Case(ii): $L_0|\psi_1\rangle ={3k\over 2}|\psi_1\rangle$}

Although the calculation is similar to case (i), the results require a
more subtle interpretation. In this case, the null vectors
are given by
\eqq
\vec {n^+} &=&
((k+1)L^+_{-2} - U_{-1}L^+_{-1} -  L_{-1} L^+_{-1})
\vec {\psi_1}\\
\vec {n^0} &=& (L^+_{-1}L^-_{-1}+L^-_{-1}L^+_{-1}-(k+1)[(k-2)U_{-1}U_{-1}-
2k(k+1)L_{-2}\cont +k(L_{-1})^2])
\vec {\psi_1}\\
\vec {n^-} &=&
((k+1)L^-_{-2} + U_{-1}L^-_{-1} - L_{-1} L^-_{-1})
\vec {\psi_1}\\
\vec {n^f} &=&
( L^-_{-2}L^+_{-1}+L^+_{-2}L^-_{-1} +(k+1)[ -L^-_{-1}U_{-1}U_{-1} +5
U_{-1}U_{-2} - 2(k+1) L_{-3}\cont
+2(k+1)  L_{-1} L_{-2} - (L_{-1})^3])
\vec {\psi_1}
\enn
The recurrence relations derived from the two charged null vectors
$\vec{n^\pm}$ are
\eqq
\alpha^p_n (\phi^+)^p_n &=& N^{p-1}(k+1)\psi^{p-1}_n+(k+1)
\sum_{x>0}L^+_{-x}\psi^p_{n-x}
-U_{-x}(\phi^+)^p_{n-x}\\
\beta^p_n (\phi^-)^p_n &=& (k+1)\psi^{p+1}_n+(k+1)
\sum_{x>0}L^-_{-x}\psi^p_{n-x}
+U_{-x}(\phi^-)^p_{n-x}\, ,
\label{eq.vectorrr3}
\enn
where
\eqq
\alpha^p_n &=& n-y+k+\Lambda-p\\
\beta^p_n &=& n-y+k-\Lambda+p\, .
\enn
As before we can derive a further relation from $\vec{n^0}$ and $\vec{n^f}$.
For brevity we only quote the relation for $n=0$ which is
\eq
\gamma^p_0\psi^p_0=N^{p-1}(\phi^-)^{p-1}_0+L^+_0 (\psi^+)^p_0
\label{eq.vectorrr4}
\en
where
\eq
\gamma^p_n=(k+1)(n-y+{3k\over 2})
[(n-y)(n-y+2k-1)+(\Lambda-p)(\Lambda-p+1)-2(k+1)h_\phi]\, .
\en
Using (\ref{eq.vectorrr0}),(\ref{eq.vectorrr3}) and (\ref{eq.vectorrr4}),
we can show that
\eq
(y-{3k\over 2})[(y-k-\Lambda+p)(y+1-k+\Lambda-p)(y-a)(y+a+1-2k)-N^{p-1}]
\psi^p_0=0
\en
Solving this equation for $y$, and calculating the fourth order Casimir
on $\psi^p_0$, we find that the following fusions are allowed;
\eq
{\psi_1}(z) \times (\Lambda,a)\to(\Lambda',a')=
(\Lambda \mp [1+k],a),(\Lambda,a),(\Lambda,a\mp [1+k])\, .
\en

Once more, the fusion rules for the field $\psi_1$ resemble the tensor
decomposition rule of the vector representation, with the important
difference that the weight lattice has been scaled by $(1+k)$. Of course, the
co-existence of lattices at two different scales occurs in the representation
theory of the standard $WA_n$ algebras, where the scales are usually labelled
$\alpha_+$ and $\alpha_-$. However, in the present case there seems to be
some problem in interpreting the first two of the above fusions.
The charge of the highest weight state
$(\Lambda',a')$ has been shifted by $\pm (1+k)$ which by assumption is
irrational and certainly non-integral. The states $\psi^p_n$ have charge
$\Lambda-p$ and thus are of the form $(L^-_0)^q|\Lambda',a'\rangle$
where $q=\Lambda'-\Lambda+p=\pm(1+k)+p$ which is non-integral. It seems
therefore that the representations corresponding to $(\Lambda',a')$ are
necessarily unbounded above and below in these cases.

(This interpretation seems sensible
for the case $\Lambda'=\Lambda-1-k$. Since in this case $\alpha^0_0$ vanishes
we can have $(\phi^+)^0_0\ne 0$ and so from (\ref{eq.vectorrr0})
$L^+_0\psi^0_0\ne 0$. In general we have that
\eq
(L^+_0)^r\psi_1(z)|\phi\rangle = z^r (\hat L^+_{-1})^r\psi^1(z)|\phi\rangle
\en
In this case the states $(L^+_{-1})^r|\psi_1\rangle$ are not generally null,
and so we have no reason to believe that the left hand side of the above
equation should vanish. Note that this is in contrast to case (i) where
$(L^+_{-1})^2|\psi_1\rangle$ is null, and in that case one sees from the
allowed fusions that indeed $(L^+_0)^2\psi^0_0$ vanishes in all cases.
However the case $\Lambda'=\Lambda+k+1$ the interpretation seems more
problematic yet. In this case $\alpha^0_0$ does not vanish and it would
seem that $L^+_0\psi^0_0$ must vanish. However treating $\psi^0_0$ as a
highest weight for $\Wb$ seems to be inconsistent with the values of
the Casimir $C$
associated with $(\Lambda',a')$, and indeed the calculation of $C$ breaks
down on $\psi^0_0$ because of the vanishing of $\beta^{-1}_0$. One
way out of this is to insist that we only fuse with representations
which are neither highest nor lowest weight, so that $p=d+{\bf Z}$,
$d\notin {\bf Z}$ in all the above expressions.)

\section{General degenerate representation theory of $\Wb$}

In this section we shall combine the information obtained
in the previous section
with our expectations based on the free field representation given in
section 3 to conjecture some results about general degenerate representations
of $\Wb$.

We begin by reconsidering the free field parametrisation of the charge and
weight of a highest weight state $|\Lambda,a\rangle$. If we write the vector
$(\Lambda,a)$ as $\beta$ then the weight of the state $|\Lambda,a\rangle$
is given by (\ref{eq.hff}) in terms of $\beta$ by
\eq
h(\Lambda,a) = {\beta \cdot (\beta- 2\rho)\over 2(k+1)}
\label{eq.hvec}
\en
where $\rho=(-1/2,k-1/2)$. We define the vector ${\tilde{\b}}=\beta-\rho$, so
that
\eq
h(\Lambda,a)={{{\tilde{\b}}} ^2-\rho^2\over 2(k+1)}\, .
\en
As we already remarked in section 2, $h$ is invariant under ${\tilde{\b}}\to
w({\tilde{\b}})$
where $w\in W$, the Weyl group of $B_2$.

Suppose that the representation built up from the primary state $|\beta\rangle$
contains some singular vector $\vec \chi$. $\vec \chi$ is also highest weight
for
$\Wb$, and we label its charge and weight as above by ${\tilde{\b}} '$. We
should like
to argue that
\eq
{\tilde{\b}} '-{\tilde{\b}}= -a_i\alpha_i
\label{eq.lattice}
\en
where $a_i$ is a positive integer and
$\alpha_i$ is a root of $B_2$. It comes as no surprise that the charge of the
primary state and its singular descendent is an integer, since the charge of
all descendents of the primary states have this property. That the value of
$a$ should differ in this way is more difficult to see.
This can be checked in all the examples
considered in the previous section. For example, $\vec \b$ has a singular state
of the form $(L^-_0)^p \vec \b$ if $a=\Lambda+k-p$. It is easy to check that
${\tilde{\b}} '-{\tilde{\b}}=(-p,p)$ in this case. Similarly, we showed that
for $a=k-n(k+1)$,
$\vec \b$ has a singular vector at level $n$ and charge $\Lambda$. In this
case,
${\tilde{\b}} '-{\tilde{\b}}=(0,-1)$.

{}Further evidence that (\ref{eq.lattice}) holds
comes from considering the screening
charge defined at the end of section 2. The standard construction of singular
vectors using a screening charge $S$ involves the expression
$(S)^n|\b'\rangle$ where $(S)^n$ is a multiple integral with suitably
defined contours, such that for certain values of $\b'$ this expression
does not vanish, but is a singular vector. Since $\Wb$ commutes with $S$, the
charge and weight of the singular vector are given by $\b'$, but given the
form of the screening charge (\ref{eq.screening}) we have
\eq
(S)^n|\b'\rangle = \{X_{-m}...\}|\b'-(n,0)\rangle
\en
where $\{X_{-m}...\}$ represent creation operators of $\Wb$. Thus we expect
from the form of the screening charge
that $\b'-\b={\tilde{\b}} '-{\tilde{\b}}=(-n,0)$ in this case.
One can generalise this argument to yield other roots of $B_2$ in
(\ref{eq.lattice}) by introducing unreduced screening charges associated with
the horizontal $su(2)$ currents.

{}From now on we shall merely assume the validity of (\ref{eq.lattice}).
The weight of $\vec \b$ and its descendent $\chi$ differ by an integer, so we
have that
\eq
2(k+1)(h'-h)={\tilde{\b}}'^2-{\tilde{\b}}^2=({\tilde{\b}}'-{\tilde{\b}})\cdot
({\tilde{\b}}'+{\tilde{\b}})=2N(k+1)
\label{eq.hdiff}
\en
If we use (\ref{eq.lattice}), and put $N=a_ib_i$ where $b_i$ is a non
negative integer, and as before $a_i$ is a positive integer,
we find that
\eq
{\tilde{\b}} \cdot \alpha_i = a_i {(\alpha_i)^2\over 2}-b_i(k+1)\, .
\label{eq.degenerate}
\en
If $\tilde{\b}$ satisfies this equation, then there is a singular vector
in the Verma module built up from $\vec {\tilde{\b}}$
with
\eq
{\tilde{\b}}'=\tilde{\b}-a_i \alpha_i
\en
at level $a_ib_i$ and with charge $\Lambda-a_i(\alpha_i\cdot \alpha_2)$.
A more convenient parametrisation of $\tilde{\b}$ is given by
\eq
{\tilde{\b}}= \sum_{i=1}^{i=2} a_i \lambda_i -(1+k)b_i \lambdav_i
\label{eq.fpar}
\en
where $\lambda_i \cdot \alpha_j=\delta_{ij}(\alpha_j)^2/2$, and
$\lambdav_i \cdot \alpha_j=\delta_{ij}$ are the fundamental weights and
coweights of $B_2$.  With this parametrisation (\ref{eq.degenerate}) is
satisfied for $\alpha_1$, $\alpha_2$, the simple roots of $B_2$.
We shall call the corresponding primary state
\eq
\pmatrix{a_1 & b_1\cr
         a_2 & b_2}
\en

Equation (\ref{eq.degenerate})
is the most important in this section. From it we shall now
write down character formulae, general formulae for the fusion of maximally
degenerate representation and give a determinant formulae for $\Wb$.

\subsection{Character formulae for representations of $\Wb$}

In the following paragraphs we give character formulae for representations
of $\Wb$. As in the rest of this paper we shall restrict ourselves to the
case that $k$ is irrational for
simplicity,
and for brevity's sake, we shall further
restrict ourselves to some interesting examples.

If $\b$ is such that (\ref{eq.degenerate}) is not satisfied for any $\alpha_i$,
then it follows that the representation built up from $\vec \b$ contains
no singular vectors and is irreducible. The character for a highest weight
representation is given by
\eqq
\chi_{(\Lambda,a)}(x,q)&=& {\rm Tr}_{(\Lambda,a)}(q^{L_0}x^{U_0})
=q^h x^\Lambda F(x,q)\\
F(x,q) &=& {1\over (1-x^{-1})}\prod_{n=1}^{\infty}
{1\over {(1-q^nx^{-1})(1-q^n)^2(1-q^n x)}}
\label{eq.chnd}
\enn

At the opposite extreme, suppose that $\b$ is such that with the
parametrisation
(\ref{eq.fpar}), $a_i, b_i$ are all positive integers.
We calculate the embedding pattern
for the singular vectors in this case by checking for descendent singular
states
using the condition
(\ref{eq.degenerate}), using (\ref{eq.lattice}) to determine the weight and
charge of the singular states so found, and then iterating the process.

\vskip 4mm

\setlength{\unitlength}{0.012500in}%
\begin{picture}(220,357)(170,465)
\thicklines
\put(345,730){\vector( 2,-1){ 86}}
\put(435,730){\vector(-2,-1){ 86}}
\put(335,730){\vector( 0,-1){ 45}}
\put(445,730){\vector( 0,-1){ 45}}
\put(345,650){\vector( 2,-1){ 86}}
\put(435,650){\vector(-2,-1){ 86}}
\put(335,650){\vector( 0,-1){ 45}}
\put(445,650){\vector( 0,-1){ 45}}
\put(380,800){\vector(-1,-1){ 40}}
\put(400,800){\vector( 1,-1){ 40}}
\put(340,575){\vector( 1,-1){ 40}}
\put(440,575){\vector(-1,-1){ 40}}
\put(300,740){\makebox(0,0)[lb]{\smash{$_{(\Lambda +a_1,h +a_1b_1}$)}}}
\put(410,740){\makebox(0,0)[lb]{\smash{$_{(\Lambda-a_2,h+a_2b_2)}$}}}
\put(225,660){\makebox(0,0)[lb]{\smash{
$_{(\Lambda-a_1-a_2,h+a_1b_1+a_2b_2+2a_1b_2)}$}}}
\put(395,660){\makebox(0,0)[lb]{\smash{
$_{(\Lambda+a_1,h+a_1b_1+a_2b_2+a_2b_1)}$}}}
\put(225,585){\makebox(0,0)[lb]{\smash{
$_{(\Lambda-a_1-a_2,h+[a_1+a_2][b_1+2b_2])}$}}}
\put(395,585){\makebox(0,0)[lb]{\smash{
$_{(\Lambda,a-2a_1-a_2,h+[2a_1+a_2][b_1+b_2])}$}}}
\put(300,510){\makebox(0,0)[lb]{\smash{
$_{(\Lambda-a_2,h+2a_1b_1+2a_2b_2+2a_1b_2+a_2b_1)}$}}}
\put(375,810){\makebox(0,0)[lb]{\smash{$_{(\Lambda,h)}$}}}
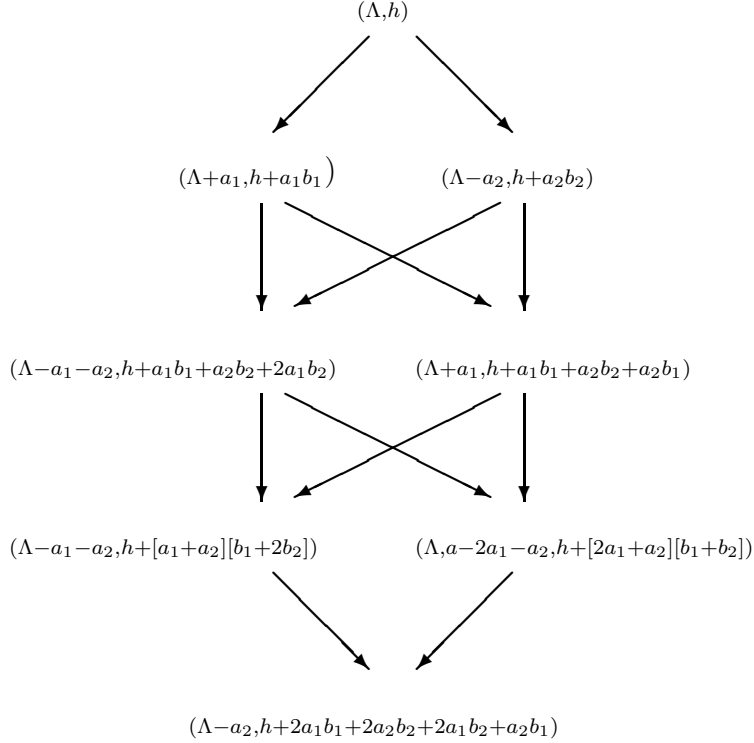
\put(260,465){\makebox(0,0)[lb]{\smash{Fig1: Null Vector embedding diagram
for integral $a_i$, $b_i$}}}
\end{picture}

\vskip 4mm

Strictly speaking if we find two singular states with the same charge and
weight
by two different paths we cannot be sure that they are the same vector, but
we shall assume this is true, relying on our experience with other W-algebras.
For $a_i,b_i$ all positive integers,
we get the embedding diagram of singular vectors given in fig. 1.
Each singular
vector is labelled in this diagram by its charge and weight, which are given
in terms of the charge and weight $(\Lambda,h)$ of the highest weight
state. The
character in this case is given by
\eqq
 &\, & [(1-q^{(2a_1+b_1)(b_1+b_2)})-x^a_1(q^{a_1b_1}-q^{a_1b_1+a_2b_2+a_2b_1})
-x^{-a_2}(q^{a_2b_2}-q^{2a_1b_1+2a_2b_2+2a_1b_2+a_2b_1})\cont
+ x^{-a_1-a_2}(q^{a_1b_1+a_2b_2+2a_1b_2}-q^{(a_1+a_2)(b_1+2b_2)}]F(x,q)
\label{eq.chcd}
\enn

Finally let us turn our attention to representations of $\Wb$ with
finite dimensional highest weight space. We need a singular state at level
zero with charge $p$ less than that of the highest weight state. This
implies that we must have $a_ib_i=0$ and $-a_i(\alpha_i\cdot \alpha_2)=p$
There are three solutions: $a_i=p$, $b_i=0$ and $\alpha_i=-\alpha_1,\alpha_2,
\alpha_4$. These correspond to the three solutions given by $N^{p-1}=0$
[{\it c.f.}(\ref{eq.normzero})]. Let us concentrate on the solution
$2\Lambda\in {\bf Z}$. These correspond to
\eq
\pmatrix{a_1 & b_1\cr
         a_2 & 0}
\label{eq.label}
\en
where $a_2=p=2\Lambda+1$. The most complicated embedding arises if we
take $a_1$ and $b_1$ also to be positive integers; this amounts to a
special case of the embedding diagram we just considered with $b_2$ set
to zero.
The expression (\ref{eq.chcd}) in this case reduces to
\eq
  [(1-x^{-2\Lambda-1})(1-q^{(2a_1+2\Lambda+1)(b_1+b_2)})
-(x^a_1-x^{-a_1-2\Lambda-1})(q^{a_1b_1}-q^{(a_1+2\Lambda+1)b_1})]F(x,q)\,.
\label{eq.chzcd}
\en
Embedding patterns for other possible values of $a_1$, $b_1$
are given in fig.2 to give the reader some idea of the different possibilities.
The diagram is labelled similarly to fig.1.
 From the above we can see that not all degenerate representations
with two independent singular vectors have the same singular vector structure.
This
is also true for standard $W$-algebras. In analogy with that case, we shall
reserve the terminology `maximally degenerate' for those representations
with the singular vector structure given in figure 1. It turns out that all of
the representations considered in the previous sections are of this type;
that is they are labelled by (\ref{eq.label}) with $a_1$, $a_2$ and $b_1$
given by positive integers. The vacuum representation,
the spinor representation and
case (i) and (ii) of the vector representation considered in the
previous section are given by
\eq
\pmatrix{1&1\cr 1&0}\,\,\,,\,\,\,
\pmatrix{1&1\cr 2&0}\,\,\,,\,\,\,
\pmatrix{2&1\cr 1&0}\,\,\,,\,\,\,
\pmatrix{1&2\cr 1&0}
\en
respectively.

\subsection{Fusion of maximally degenerate representations}

The examples in the preceding section suggest the following generalisations
of the fusion rules proved there for primary fields with $b_1=1$.
In this case we may write
\eq
\beta=(a_1-1)\lambda_1+(a_2-1)\lambda_2
\label{eq.extra}
\en
\vskip 4mm
\setlength{\unitlength}{0.012500in}%
\begin{picture}(575,442)(130,380)
\thicklines
\put(130,730){\vector( 2,-1){ 86}}
\put(220,730){\vector(-2,-1){ 86}}
\put(120,730){\vector( 0,-1){ 45}}
\put(230,730){\vector( 0,-1){ 45}}
\put(165,800){\vector(-1,-1){ 40}}
\put(185,800){\vector( 1,-1){ 40}}
\put(125,575){\vector( 1,-1){ 40}}
\put(225,575){\vector(-1,-1){ 40}}
\put(130,650){\vector( 2,-1){ 86}}
\put(220,650){\vector(-2,-1){ 86}}
\put(120,650){\vector( 0,-1){ 45}}
\put(230,650){\vector( 0,-1){ 45}}
\put(360,725){\vector( 2,-1){ 86}}
\put(450,725){\vector(-2,-1){ 86}}
\put(350,725){\vector( 0,-1){ 45}}
\put(460,725){\vector( 0,-1){ 45}}
\put(355,650){\vector( 1,-1){ 40}}
\put(455,650){\vector(-1,-1){ 40}}
\put(295,445){\vector( 1,-1){ 40}}
\put(395,445){\vector(-1,-1){ 40}}
\put(335,515){\vector(-1,-1){ 40}}
\put(355,515){\vector( 1,-1){ 40}}
\put(395,795){\vector(-1,-1){ 40}}
\put(415,795){\vector( 1,-1){ 40}}
\put(515,500){\vector( 0,-1){ 60}}
\put(620,500){\vector( 0,-1){ 60}}
\put(585,790){\vector(-1,-1){ 40}}
\put(605,790){\vector( 1,-1){ 40}}
\put(545,720){\vector( 1,-1){ 40}}
\put(645,720){\vector(-1,-1){ 40}}
\put( 80,785){\makebox(0,0)[lb]{\smash{{\bf (i)}}}}
\put(305,780){\makebox(0,0)[lb]{\smash{{\bf (ii)}}}}
\put(500,780){\makebox(0,0)[lb]{\smash{{\bf (iii)}}}}
\put(255,520){\makebox(0,0)[lb]{\smash{{\bf (iv)}}}}
\put(460,540){\makebox(0,0)[lb]{\smash{{\bf (v)}}}}
\put(560,540){\makebox(0,0)[lb]{\smash{{\bf (vi)}}}}
\put(160,810){\makebox(0,0)[lb]{\smash{$_{(\Lambda,h)}$}}}
\put(200,740){\makebox(0,0)[lb]{\smash{$_{(-\Lambda-1,h)}$}}}
\put( 85,740){\makebox(0,0)[lb]{\smash{$_{(\Lambda+a_1,h+a_1b_1)}$}}}
\put( 70,660){\makebox(0,0)[lb]{\smash{$_{(-\Lambda-a_1-1,h+a_1b_1)}$}}}
\put(185,660){\makebox(0,0)[lb]{\smash{$_{(\Lambda+a_1,h+[a_1+a_2]b_1)}$}}}
\put( 75,580){\makebox(0,0)[lb]{\smash{$_{(-\Lambda-a_1-1,h+[a_1+a_2]b_1)}$}}}
\put(215,580){\makebox(0,0)[lb]{\smash{$_{(\Lambda,h+[2a_1+a_2]b_1)}$}}}
\put(130,510){\makebox(0,0)[lb]{\smash{$_{(-\Lambda-1,h+[2a_1+a_2]b_1)}$}}}
\put(390,810){\makebox(0,0)[lb]{\smash{$_{(\Lambda,h)}$}}}
\put(340,735){\makebox(0,0)[lb]{\smash{$_{(-\Lambda-1,h)}$}}}
\put(415,735){\makebox(0,0)[lb]{\smash{$_{(\Lambda+a_1,h+[a_1+a_2]b_1)}$}}}
\put(350,590){\makebox(0,0)[lb]{\smash{$_{(-\Lambda-1,h+[2a_1+a_2]b_1)}$}}}
\put(440,660){\makebox(0,0)[lb]{\smash{$_{(\Lambda,h+[2a_1+a_2]b_1)}$}}}
\put(300,660){\makebox(0,0)[lb]{\smash{$
_{(-\Lambda-1-a_1,h+[a_1+a_2]b_1)}$}}}
\put(580,805){\makebox(0,0)[lb]{\smash{$_{(\Lambda,h)}$}}}
\put(580,805){\makebox(0,0)[lb]{\smash{$_{(\Lambda,h)}$}}}
\put(330,530){\makebox(0,0)[lb]{\smash{$_{(\Lambda,h)}$}}}
\put(505,515){\makebox(0,0)[lb]{\smash{$_{(\Lambda,h)}$}}}
\put(610,515){\makebox(0,0)[lb]{\smash{$_{(\Lambda,h)}$}}}
\put(275,455){\makebox(0,0)[lb]{\smash{$_{(-\Lambda-1,h)}$}}}
\put(495,415){\makebox(0,0)[lb]{\smash{$_{(-\Lambda-1,h)}$}}}
\put(595,415){\makebox(0,0)[lb]{\smash{$_{(-\Lambda-1,h)}$}}}
\put(350,455){\makebox(0,0)[lb]{\smash{$_{(\Lambda+a_1,h+[a_1+a_2]b_1)}$}}}
\put(285,380){\makebox(0,0)[lb]{\smash{
$_{(-\Lambda-1-a_1,h+[a_1+a_2]b_1)}$}}}
\put(440,660){\makebox(0,0)[lb]{\smash{$_{(\Lambda,h+[2a_1+a_2]b_1)}$}}}
\put(580,730){\makebox(0,0)[lb]{\smash{$_{(\Lambda,h+[2a_1+a_2]b_1)}$}}}
\put(520,730){\makebox(0,0)[lb]{\smash{$_{(-\Lambda-1,h)}$}}}
\put(545,655){\makebox(0,0)[lb]{\smash{$_{(-\Lambda-1,h+[2a_1+a_2]b_1)}$}}}
\end{picture}
\vskip 4mm
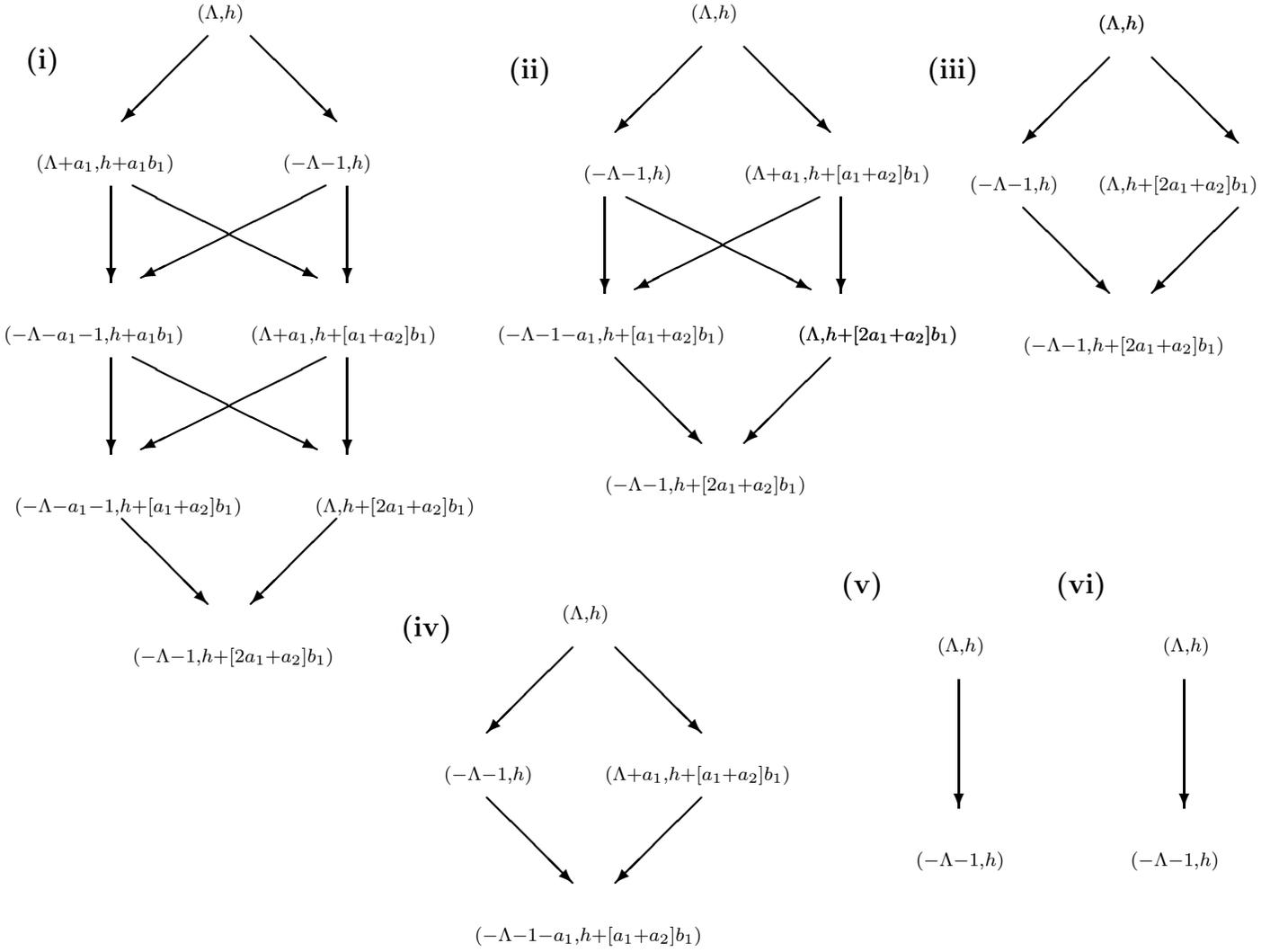
\centerline{Fig2: Null vector embedding diagram for $b_2=0$,$a_2,b_1\in Z_+$
and}
\centerline{$a_1\in Z$ with (i) $a_1>0$ (ii) $-a_2/2<a_1\le 0$ (iii)
$-a_2<a_1\le -a_2/2$ (v) $a_1\le -a_2$ or}
\centerline{$a_1\in Z+1/2$ and (iv)$-a_2/2<a_1$ (vi) $a_1 \le  a_2/2$}
\vskip 5mm

We can interpret $\beta$
as a highest weight of a (finite) $B_2$ representation.
A natural set of fusion rules for $\phi_\beta$ with a non-degenerate primary
field are
\eq
\phi_{\beta}\times \phi_{\beta'}\to\phi_{\beta'+\lambda}
\en
where $\lambda$ ranges over the different weights of the representation
with highest weight $\beta$, so that the number of possible fusions is simply
the dimension of this representation. This is consistent with
what we found for the spinor
representation and case (i) of the vector representation in the previous
section.

Some support for this conjecture can be found by considering the number of
constraints on the operator product expansion that the null vectors of
$\phi_{\beta}$ imply \cite{BWat3}. If we let $B(n)$ be the difference between
the number of independent
operator product expansions of the form (\ref{eq.opebasis}) and the
number of independent constraints arising from null vectors up to level $n$,
then we find that
\eqq
\sum_{n\ge 0} B(n)q^n &=&
\lim_{x\to 1}x^{-\Lambda} q^{-h}{\chi_{\Lambda_1}(x,q)\over \chi_{vac}(x,q)}\\
&=& {a_2(1-q^{2a_1+a_2})-(2a_1+a_2)(q^{a_1}-q^{a_1+a_2})\over (1-q)^3}
\enn
This turns out to be a finite polynomial in $q$ of order
$2a_1+a_2-3$. So for $n>2a_1+a_2-3$, the number of constraints from
null vectors exactly matches the number of basis vectors of OPE's, so
adding weight to the claim that we can always solve for the fusion of
such representations. Moreover it can be argued that the number of
independent fusions is given by
\eq
\sum_{n\ge 0} B(n)=
\lim_{x,q\to 1} x^{-\Lambda}q^{-h}{\chi_{\Lambda_1}(x,q)\over \chi_{vac}(x,q)}
={a_1 a_2(a_1+a_2)(2a_1+a_2)\over 6}
\en
which is the dimension of a representation of $B_2$ with highest weight
$\beta$.

Unfortunately, these arguments do not seem to make much sense in the
case where $b_1\ne 1$. For example, let us consider taking $a_1=a_2=1$.
In this case we may write
\eq
\beta = (1+k)[(b_1-1)\lambdav_1+b_2\lambdav_2]
\en
and we might guess that we should interpret $\beta$ as a highest weight for
the Lie algebra with $\alpha^\vee_i$ as roots.
This is $C_2$ which is isomorphic to $B_2$ again, but with simple roots
$(-1,1)$, $(2,0)$. However,  with this interpretation, case (ii) of the vector
representation considered in section 5 corresponds to $b_1=2$, $b_2=0$ which
in turn
should correspond to the spinor representation of $C_2$.
This would only produce four of the five fusions found in the
previous section. The conditions established there however were only necessary,
not sufficient, so it is not inconceivable that the extra fusion (which
corresponds to no shift in $(\Lambda,a)$) is disallowed by conditions coming
from higher level null vectors. Worse still, the calculation used above to
calculate the number of fusions when $b_1=1$ is of no help here.
It gives the answer $b_1^3$ for the
number of possible fusions of the field corresponding to $\pmatrix{
1 & b_1\cr 1 & 0}$, which disagrees with both the above answers for
$b_1=2$. Clearly this deserves a more careful treatment.
Nonetheless, the results we have obtained do not seem incompatible the
elegant conjecture that if
\eq
\beta=(a_1-1)\lambda_1+(a_2-1)\lambda_2+(1+k)[(b_1-1)\lambdav_1+b_2\lambdav_2]
\en
then the fusion of the corresponding field $\phi_\beta$ is given by
\eq
\phi_{\beta}\times \phi_{\beta'}\to\phi_{\beta'+\lambda+(1+k)\lambda'}
\en
where $\lambda$ is a weight of the $B_2$ representation with highest weight
$(a_2-1)\lambda_1+(a_2-1)\lambda_2$ and $\lambda'$ is the weight of the $C_2$
representation with highest weight $(b_1-1)\lambdav_1+b_2\lambdav_2$.

\subsection{Determinant formula for $\Wb$}

It is straightforward to see what the generalisation of the determinant
formulae for standard $W$-algebras \cite{WDet} is,
given (\ref{eq.lattice}). It is
convenient to Taylor expand $F(x,q)$ as follows
\eq
F(x,q)=\sum_{a,b}U_{cd}q^cx^d
\en
If we consider the space of descendent states of
$\vec \beta=|\Lambda,a\rangle$ with
charge $\Lambda+M$, and at level $N$, then we expect that the determinant
of the matrix of inner products of states in this space is given by
\eq
{\rm det}_{MN}(\beta)
=C \prod_{j=0}^N \prod_{\{r>0,s\ge 0:rs=j\}}\prod_{\alpha\in\Delta}
( {\tilde{\b}} \cdot \alpha_i - r{(\alpha_i)^2\over 2}+s(k+1))^{U_{cd}}
\label{eq.dform}
\en
where $C$ is a constant, and $c=N-rs$, $d=M+r(\alpha \cdot \alpha_2)$.

%We can support the claim that this is indeed the determinant formula for
%$\Wb$ by calculating the order of
%$det_{MN}$ as a polynomial in $\tilde{\beta}$.
%(PROOF)

\section{Conclusions}

In this paper we have constructed a $W$-algebra of remarkable simplicity
which we believe has most of the features of $W$-algebras
arising from Drinfel'd-Sokolov reductions. This makes it a useful laboratory
for studying many of the unresolved issues concerning
such algebras. In particular the representation theory and
modular properties of the characters are questions that need
to be addressed in order to construct the Hilbert space of the
corresponding statistical systems.

Some steps towards understanding the representation theory have been made
in this work. The zero mode algebra was found to be an interesting
polynomial deformation of the zero-grade algebra $g_0$ ($su(2)\times u(1)$ in
this case). We expect this to hold in the general case since we know that
in the $c\to \infty$ limit, the zero mode algebra coincides with $g_0$
\cite{BWat1}.
Many of the formulae which were derived by free field techniques in the
standard reduction case can be taken over simply by adjusting the `shift
vector' $\rho$.
Maximally degenerate representations could be labelled by four
integers
\eq
\pmatrix{a_1 & b_1\cr a_2 & b_2}
\en
exactly as in the case of the standard reduction {\cite{KWat2}).
However, only some of these representations have a finite highest
weight space $V_0$. We considered the case where $b_2=0$ which
corresponds to half-integrally charged state.  Interestingly, only a
subset of these representations, corresponding to those with $b_1=1$
close under fusion to give states with finite highest weight spaces.

There are many questions that need to be addressed by future work. We have
not had time to consider rational values of the central charge $c$ (or
rather the parameter $k$). We expect that as in the standard case, for
rational $k$ we will have a periodic identification of completely
degenerate representations \cite{BPZ},
and this will lead to some truncation in the
fusion rules. It should also be fairly straightforward to derive character
formulae for these representations by analogy to the standard case, and
to find how the characters transform under the modular group. It will then
be possible to write down modular invariant partition functions for $\Wb$
models. Also the issue of unitarity of representations has not been touched
in this paper. A necessary condition for unitarity is that the
the central charge (\ref{eq.cch}) and the parameter $k$ be simultaneously
positive. This only occurs for $4-\sqrt{10}<6k<4+\sqrt{10}$, a fairly
restrictive range. This ties in with the prejudice
that the standard reduction of simply-laced
algebras are rather special in that they each admit an infinite
sequence of unitary models. For more general reductions, it seems rather
harder to find unitary examples.
Many of the formulae considered in the paper seem to have elegant
generalisations for all reductions, and give important clues as to how a
unified treatment of the general case would look. In particular, the free
field parametrisation of the weights of degenerate fields involves both
weights and coweights, as we might have anticipated
from the duality arguments of \cite{KWat1}.
By comparing the primary field content and fusion
rules that we obtain, we can compare the conformal field theories constructed
by
reductions and by the coset method, and this may give us some prescription
for going between the two (which already exists for the theories with
symmetry algebra $W_g$ if $g$ is simply-laced \cite {BBSS}).
This would provide us with two descriptions of
the same conformal field theory, one in which the properties of the chiral
algebra were transparent, and the other in which the modular properties of
characters were clear.
We intend to pursue this further elsewhere. Finally,
$\Wb$ is such a simple algebra that it would be surprising if the
corresponding conformal models, or their integrable perturbations did not have
some physical significance.

\noindent
{\bf Acknowledgements}

It gives me great pleasure to thank G.M.T.Watts for many stimulating
conversations. I should also like to
thank Adrian Kent, Koos de Vos and Cosmas Zachos for discussions.
Part of this work was completed at the University of Chicago and was
supported by the following grants:
U.S. DOE grant  DEFG02-90-ER-40560 and NSF grant PHY900036

\section{Appendix}

In this appendix we demonstrate that the screening charge $S$ given
by (\ref{eq.screening})
commutes with the generators of $\Wb$. By construction $S$ is the integral
of a current of
conformal weight one and zero charge with respect to $U_0$. This
ensures that it commutes with both $L(z)$ and $U(z)$. We now show that it
commutes with the expressions for $L^\pm(z)$ given by (\ref{eq.freecharge}).

The operator product of $L^+$ with the screening current $s$ is of the form
\eq
L^+(z)s(\zeta)\equiv {{\hat L}^+_0 s(\zeta)\over (z-\zeta)^2}+
{{\hat L}^+_{-1} s(\zeta)\over (z-\zeta)}
\en
For $L^+(z)$ to commute with $S=\oint dz s(z)$, the right hand side of
this equation must be a total derivative in $\zeta$, that is
\eq
L^+_{-1}|s\rangle=L_{-1}L^+_0|s\rangle
\label{eq.chargecommute}
\en

Substituting in the free field representation (\ref{eq.freecharge})
for $L^+(z)$, we see that
\eqq
L^+_{-1}|s\rangle &=& [j^+_{-1}j^1_0+j^1_{-1}j^+_0+r^+(j^2_{-1}j^+_0
                       +j^+_{-1}j^2_0)]|s\rangle\\
                  &=& (j^1_{-1}+r^+j^2_{-1})|s^+\rangle+r^+\alpha|s\rangle\\
L^+_{0}|s\rangle &=& [j^+_{0}j^1_0+r^+
                       +j^+_{0}j^2_0-s^+j^+_0]|s\rangle\\
                  &=& (r^+\alpha-s^+)|s^+\rangle
\enn
where $|s^+\rangle=j^+_0|s\rangle$.
Also using (\ref{eq.impvir}) we have that
\eqq
L_{-1}|s^+\rangle &=& [{\cal L}_{-1}+j^2_{-1}j^2_0]|s^+\rangle\\
                  &=& {1\over k+1}(j^1_{-1}|s^+\rangle+j^+_{-1}|s\rangle)
                       +\alpha j^2_{-1}|s^+\rangle
\enn
Subsituting in $r^+=-\sqrt{(k+1)},s^+=-k$, one easily reads off that
(\ref{eq.chargecommute}) is satisfied for $\alpha=-1/\sqrt{(k+1)}$, but not
for $\alpha=2k/\sqrt{(k+1)}$. The same result holds if we consider $L^-(z)$
instead.

\end{document}